\documentclass[reprint,linenumber, amsmath amssymb, titlepage, aps, pra, superscriptaddress, raggedbottom, nofootinbib, floatfix, citeautoscript]{revtex4-2}
\usepackage{silence}
\usepackage[utf8]{inputenc}
\usepackage[english]{babel}
\usepackage{hyperref}
\usepackage{bm}
\usepackage{hyperref}
\usepackage{float}
\usepackage[newcommands]{ragged2e}
\usepackage{physics}
\usepackage{tabularx}
\usepackage{mathtools}
\usepackage[raggedright]{titlesec}
\usepackage[toc,page,header]{appendix}
\usepackage{nicefrac}
\usepackage{sidecap}
\usepackage{multibib}
\usepackage[dvipsnames]{xcolor}
\usepackage{booktabs}
\usepackage{float}
\usepackage{graphicx}
\usepackage[capitalise]{cleveref}
\usepackage[version=4]{mhchem}
\usepackage{tabularx}
\usepackage{enumitem}
\usepackage[textsize=scriptsize, color=red!90, textcolor=white]{todonotes}
\sidecaptionvpos{figure}{c}
\usepackage[per-mode=power, 
exponent-product = \times, 
inter-unit-product = \ensuremath{{\cdot}}, 
tight-spacing = true,
number-unit-product=~,
parse-numbers=false,]{siunitx}
\AtBeginDocument{\RenewCommandCopy\qty\SI}
\ExplSyntaxOn
\msg_redirect_name:nnn { siunitx } { physics-pkg } { none }
\ExplSyntaxOff
\hypersetup{colorlinks=true,  linkcolor=blue, 
  citecolor=blue, filecolor=blue,  
urlcolor=blue}
\usepackage{caption}
\usepackage{minitoc}
\newcommand{\circled}[1]{\raisebox{.5pt}{\textcircled{\raisebox{-.9pt} {#1}}}}
\titleformat*{\section}{\bfseries}
\titleformat*{\subsection}{\bfseries}
\titlespacing{\section}{0em}{1em}{0.25em}
\renewcommand{\textcite}[1]{\citenum{#1}}
\newcolumntype{Y}{>{\centering\arraybackslash}X}
\crefname{appendix}{Supplementary Information}{Supplementary Information}
\Crefname{appendix}{Supplementary Information}{Supplementary Information}
\let\origaddcontentsline\addcontentsline
\newcommand{\disabletocentries}{%
  \renewcommand{\addcontentsline}[3]{}%
}
\newcommand{\enabletocentries}{%
  \let\addcontentsline\origaddcontentsline
}
\setcitestyle{super}
\DeclareSIUnit[qualifier-mode = combine]{\dBm}{\deci\bel\of{m}}

\usepackage{ulem}
\usepackage{transparent}

\begin{document}
\disabletocentries
\author{Gr\'egory~Moille}%
\email{gmoille@umd.edu}
\affiliation{Joint Quantum Institute, NIST/University of Maryland, College Park, USA}
\affiliation{Microsystems and Nanotechnology Division, National Institute of Standards and Technology, Gaithersburg, USA}
\author{Pradyoth~Shandilya}%
\affiliation{University of Maryland Baltimore County, Baltimore, MD, USA}
\author{Jordan~Stone}%
\affiliation{Microsystems and Nanotechnology Division, National Institute of Standards and Technology, Gaithersburg, USA} %
\author{River~Beard}%
\affiliation{AV Incorporated, Albuquerque, New Mexico, USA}
\author{Shao-Chien~Ou}%
\affiliation{Joint Quantum Institute, NIST/University of Maryland, College Park, USA}
\affiliation{Microsystems and Nanotechnology Division, National Institute of Standards and Technology, Gaithersburg, USA}
\author{Zongda~Li}
\affiliation{Department of Physics, University of Auckland, Auckland 1010, New Zealand}
\affiliation{The Dodd-Walls Centre for Photonic and Quantum Technologies, New Zealand}
\author{Mark~Harrington}
\affiliation{Department of Electrical and Computer Engineering, University of California Santa Barbara, Santa Barbara, CA, USA}
\author{Kaikai~Liu}
\affiliation{Department of Electrical and Computer Engineering, University of California Santa Barbara, Santa Barbara, CA, USA}
\author{Robert~Rockmore}
\affiliation{Air Force Research Laboratory, Space Vehicles Directorate, Kirtland Air Force Base 87117, USA}
\author{Curtis~R.~Menyuk}%
\affiliation{University of Maryland Baltimore County, Baltimore, MD, USA}
\author{Daniel~J.~Blumenthal}
\affiliation{Department of Electrical and Computer Engineering, University of California Santa Barbara, Santa Barbara, CA, USA}
\author{Sean~P.~Krzyzewski}
\affiliation{Air Force Research Laboratory, Space Vehicles Directorate, Kirtland Air Force Base 87117, USA}
\author{Miro Erkintalo}%
\affiliation{Department of Physics, University of Auckland, Auckland 1010, New Zealand}
\affiliation{The Dodd-Walls Centre for Photonic and Quantum Technologies, New Zealand}
\author{Kartik~Srinivasan}
\email{kartik.srinivasan@nist.gov}
\affiliation{Joint Quantum Institute, NIST/University of Maryland, College Park, USA}
\affiliation{Microsystems and Nanotechnology Division, National Institute of Standards and Technology, Gaithersburg, USA}

\date{\today}

\newcommand{\mytitle}{Self-aligned optical microcomb emerging between octave separated lasers} 

\title{\mytitle}

\begin{abstract}
  Optical frequency combs (OFCs) are frequency rulers essential for precision metrology, next generation navigation, and testing of fundamental physics~\cite{DiddamsScience2020}. %
  Despite intense efforts, chip-integrated OFCs remain laboratory-bound, unable to fulfill their promise of compact and cost-effective deployment. %
  While improvements in fabrication and integration are important, a conceptual limitation has  stymied progress:
  on-chip OFC architectures have aimed to miniaturize their table-top counterparts and relied on cascading outward from  a single pump~\cite{KippenbergScience2018,LiOptica2017}. %
  In integrated platforms, this approach does not readily generate the strong, low-noise, octave-spaced signals crucial for robust zero-frequency offset detection~\cite{DrakeNat.Photonics2020}. %
  Here, we overcome this limitation via an architectural inversion where an optical microcomb fills the spectrum between two octave-separated pumps. %
  The two pumps generate a parametrically driven cavity soliton (PDCS)~\cite{EnglebertNat.Photon.2021} in an integrated $\chi^{(3)}$ resonator,~\cite{MoilleNat.Photon.2024} which robustly self-aligns to  both pumps across multiple foundry-fabricated devices and operating configurations. %
  This produces a single octave-spanning comb from telecom to visible wavelengths, whose zero-frequency offset is defined by the pumps' harmonic offset and can therefore be reliably detected and stabilized. %
  We showcase our platform's capabilities by executing all of the three core tasks of OFC metrology: optical frequency synthesis, low-noise millimeter-wave generation, and integrated optical clock readout, using the same self-aligned microcomb with only its input locks changed.
\end{abstract}
\maketitle

\begin{figure*}[!t]
  \centering
  \includegraphics{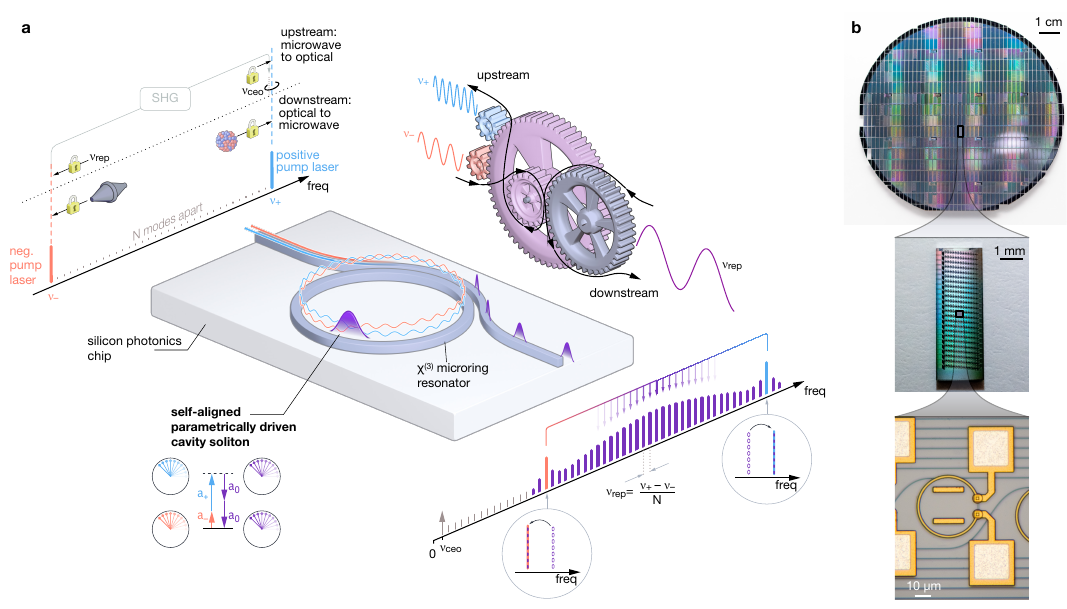}
  \caption{\label{fig:1} \textbf{%
      Concept of self-aligned parametrically driven cavity soliton (PDCS) microcomb generation with octave-separated lasers%
    }---
    \textbf{a,} Schematic of dual-pump PDCS generation. %
    Two pump lasers (with frequencies $\nu_-$ and $\nu_+\approx2\nu_-$) are injected into a $\chi^{(3)}$ microring resonator where they interact nonlinearly to generate a PDCS. %
    By minimizing their phase difference, the Kerr nonlinearity phase-locks all components (lower left inset), leading to the PDCS self-synchronizing to both of the pump fields. %
    The output microcomb spectrally fills the gap between the two pump lasers, with both capturing their nearest PDCS comb teeth to form a unique, single frequency grid. %
    The repetition rate thus becomes deterministic as $\nu_\mathrm{rep}^\mathrm{pdcs} = (\nu_+ - \nu_-)/N$, where $\nu_+$, $\nu_-$, and $N$ are all known. %
    Moreover, the quasi-harmonic pump configuration enables carrier-envelope offset (CEO) definition ($\nu_{\mathrm{ceo}}$) and locking prior to comb generation, for complete system self-stabilization, which in turn enables straightforward interfacing with locked lasers and microwave sources for core OFC tasks in both the microwave and optical domains. %
    \textbf{b,} Photographs (top and center) and micrograph (bottom) of the \qty{100}{mm} foundry-fabricated silicon nitride wafer (see Methods) in which our new deployable microcomb architecture is demonstrated.
  }
\end{figure*}

Ultra-precise optical frequency measurement and stabilization enables positioning, navigation, and timing beyond GPS~\cite{LombardiIEEEInstrum.Meas.Mag.2012}, quantum computing~\cite{FinkelsteinNature2024}, quantum sensing~\cite{YePhys.Rev.Lett.2024}, fundamental physics measurements~\cite{DereviankoQuantumSci.Technol.2022}, and may redefine the second~\cite{DimarcqMetrologia2024}. %
Optical frequency combs (OFCs)---optical frequency rulers with evenly spaced markers and a known zero-frequency offset~\cite{DiddamsScience2020}---are a key enabling technology for these applications, providing a coherent optical to microwave two-way link for ultra-high precision optical frequency measurement~\cite{MetcalfOptica2019}, synthesis~\cite{HolzwarthPhys.Rev.Lett.2000}, and ultra-low noise microwave generation~\cite{FortierNaturePhoton2011}. %
Reducing the size, weight, power, and cost (SWaP-C) of OFCs through mass-scale fabrication can enable field deployment and lab-to-consumer transition along with the critical technologies they empower. %
This vision has driven significant efforts to generate OFCs in integrated photonics platforms.

Miniaturizing from the table top to the chip-scale, integrated OFCs also rely on using the second ($\chi^{(2)}$) or third ($\chi^{(3)}$) order optical nonlinearity to spectrally broaden a narrowband input to obtain evenly spaced comb teeth spanning an octave,~\cite{BruchNat.Photonics2021,WuNat.Photon.2024,DelHayeNaturePhoton2016,HicksteinPhys.Rev.Appl.2017} thus enabling $\nu-2\nu$ (also referred to as $f-2f$) nonlinear interferometry to measure and stabilize the comb's carrier-envelope offset (CEO).~\cite{TelleApplPhysB1999,JonesScience2000} %
Dissipative Kerr solitons (DKSs) in on-chip microresonators~\cite{KippenbergScience2018} constitute a particular system of interest that leverages the photonic material's $\chi^{(3)}$ nonlinearity as an ultra-broadband gain medium. %
Generated from a continuous wave (CW) pump laser, chip-integrated DKS combs can span an octave~\cite{LiOptica2017} and are compatible with mass-scale foundry fabrication~\cite{LiuNatCommun2021,OuOpt.Lett.OL2025} and low power consumption~\cite{SternNature2018}. %
However, while DKS combs have proven their effectiveness in numerous applications~\cite{KippenbergScience2018,DiddamsScience2020,SunAdv.Opt.Photon.2023}, their outward cascading from a low-power seed pump laser leads to fundamental metrology limitations: Octave-spaced signals at the comb edges are typically very weak and noisy (due to thermo-refractive fluctuations~\cite{DrakeNat.Photonics2020}), which critically hinders the detection and stabilization of the CEO. %
These limitations have restricted precise predictability of the DKS repetition rate,~\cite{LucasPhys.Rev.A2017} and direct CEO detection and locking, to a small handful of laboratory demonstrations~\cite{BraschLightSciAppl2017,SpencerNature2018,NewmanOptica2019,DrakeNat.Photonics2020,WuOptica2023,MoilleOptica2025,WuNat.Photon.2025} that  involve complex apparatuses and/or additional helper lasers, limiting deployable integrated frequency metrology. %

\indent In this work, we demonstrate a new paradigm for integrated OFC generation and self-referencing. %
Fundamentally departing from conventional comb generation architectures that rely on cascading outwards from a single pump, our concept is underpinned by recently-discovered parametrically driven cavity solitons (PDCSs)\cite{EnglebertNat.Photon.2021} that can manifest themselves in $\chi^{(3)}$ resonators driven by two CW pump lasers,\cite{MoilleNat.Photon.2024} corresponding to frequency combs %
 that are centred in between the two pumps. %
Here, we report on two transformative advances in integrated $\chi^{(3)}$ PDCS generation that underline their direct potential for comb metrology: (i) the ability to generate on-chip PDCS combs with two octave-separated pumps, and (ii) the possibility that a generated PDCS comb can nonlinearly align itself with the pumps,  resulting in a single frequency grid across the entire pump span. %
These advances are robust across multiple foundry-fabricated devices and pump configurations, and produce a single octave-spanning OFC from telecom to visible wavelengths with intrinsic two-point microcomb stabilization control. %
Crucially, because the octave-separated pumps are both comb lines (when operating in the self-alignment regime), the CEO of the subsequently generated PDCS is \textit{a priori} determined before the chip from the harmonic offset between the pumps, enabling flexible interfacing with microwave and optical frequency references. %
We show that by simply changing how the input lasers are locked, the same self-aligned PDCS can perform the three core tasks of self-referenced OFCs: synthesizing optical frequencies from known microwave frequencies, generating low-noise micro/millimeter waves from low-noise lasers, and generating a stable microwave timing signal from an atomically-referenced laser.

\section*{Concept and experimental demonstration}
Unlike a conventional DKS driven by one pump, our  concept uses two pumps at frequencies $\nu_\pm$ and azimuthal modes $n_\pm$ (\cref{fig:1}a) driving a chip-integrated $\chi^{(3)}$ microring cavity. %
When the degenerate four-wave mixing, $\nu_+ + \nu_- = 2\nu_0$, is phase- and frequency-matched, and $\nu_0=(\nu_++\nu_-)/2$ experiences anomalous dispersion, the resonator sustains a broadband PDCS around $\nu_0$~\cite{MoilleNat.Photon.2024}.  %

Inside the resonator, the PDCS nonlinearly interacts with the fields at the pump frequencies $\nu_\pm$ to produce frequency combs around those pumps. %
The Kerr-nonlinearity locks the group velocities of the PDCS and pump combs, yielding identical repetition rates (comb line spacings). %
However, the fields at $\nu_\pm$ and $\nu_0$ generally have different phase velocities, giving the combs different CEO frequencies~\cite{MenyukOpt.ExpressOE2025}. %
The output therefore consists of three frequency grids: the central PDCS and one around each pump. %
All share $\nu_\mathrm{rep}$, but the $\nu_-$ and $\nu_+$ grids are offset from the PDCS by $-\Delta\nu_\mathrm{ceo}$ and $+\Delta\nu_\mathrm{ceo}$, respectively. %

The metrological uses of PDCS combs may appear limited because they do not generally form a single frequency grid, instead exhibiting a new degree of freedom, $\Delta\nu_\mathrm{ceo}$. %
Linear theory sets this offset through resonator dispersion and pump detunings~\cite{MoilleNat.Photon.2024}. %
Although $\Delta\nu_\mathrm{ceo}=0$ would be theoretically possible, reaching and maintaining  it would require judicious dispersion engineering and pump-detuning stabilization. %
Remarkably, however, simulations show that this expectation is inaccurate at small $\Delta\nu_\mathrm{ceo}$. %
Instead, Kerr nonlinearity self-aligns the PDCS to the pumps, locking $\Delta\nu_\mathrm{ceo}=0$ over a broad parameter range, including pump frequencies (\cref{fig:1}a and Supplementary Information S.1). %
  Although this self-alignment is conceptually related to Kerr-induced synchronization (KIS)\cite{MoilleNature2023}, where a single DKS comb tooth is captured by a weak externally injected reference laser, the two phenomena differ fundamentally. %
    In KIS, the reference weakly perturbs a DKS and plays no role in the soliton's energy balance. %
  In  contrast, in the self-alignment regime described here, the two pump fields simultaneously synchronize and generate the PDCS through a fully non-perturbative interaction, essentially giving rise to a new self-aligned nonlinear state.

The possibility of PDCS self-alignment has significant ramifications for on-chip metrology applications if the two pumps generating and capturing the PDCS are to be quasi-harmonically related, i.e., $\nu_+\approx 2\nu_-$ and $n_+ = 2 n_-$. %
Because both pumps become part of the PDCS comb, they satisfy $\nu_+ = 2n_-\nu_\mathrm{rep}^\mathrm{pdcs} + \nu_\mathrm{ceo}^\mathrm{pdcs}$ and $\nu_- = n_-\nu_\mathrm{rep}^\mathrm{pdcs} + \nu_\mathrm{ceo}^\mathrm{pdcs}$. %
The CEO can therefore be readily measured and stabilized before the chip from the strong beat between one of the pumps at $\nu_+$ and the other doubled pump at $2\nu_-$. %
Thus, one can choose a standard two-point locking architecture~\cite{PappOptica2014,KudelinNature2024,Diakonov2025} with locked lasers, or set the CEO frequency for complete self-referencing while keeping another degree of freedom for repetition rate stabilization or optical locking to cavity references or atomic clocks.

\begin{figure*}[!t]
  \centering
  \includegraphics{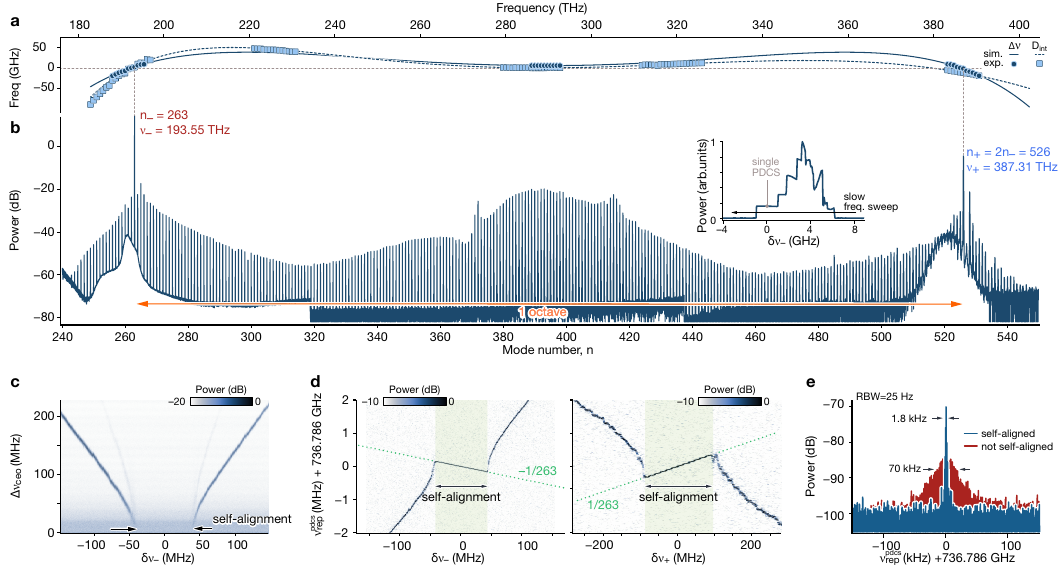}
  \caption{\label{fig:2} \textbf{Demonstration of self-aligned octave-spanning PDCS microcomb}---
    \textbf{a,} Dispersion engineering for dual-pump PDCS generation (\qty{RW = 850}{\nm}). %
    OPO frequency mismatch $\Delta\nu$ (circles: measured, solid line: simulation) when pumped at $n_- = 263$ (\qty{\nu_- = 193.55}{\THz}) and $n_+ = 2n_- = 526$ (\qty{\nu_+ = 387.31}{\THz}) is minimized at $2n_0 = n_- + n_+ + 1$, enabling PDCS generation. %
    PDCS integrated dispersion $D_{\mathrm{int}}(n-n_0)$ (squares: measured, dashed line: simulation) exhibits zero-crossing at both pumps, minimizing their phase offset with respect to the PDCS to enable efficient self-alignment. %
    \textbf{b,} Octave-spanning PDCS spectrum demonstrating a gap-free, \qty{220}{\THz} span single frequency grid with all comb teeth above \qty{-60}{\dB}. %
    The power is referenced to \qty{1}{\mW}, i.e. equivalent to \qty{}{\dBm}. %
    Inset: adiabatic PDCS access via pump tuning. %
    \textbf{c} Spectrogram from the homodyne detection of the frequency offset $\Delta\nu_\mathrm{ceo}$ between the PDCS and $\nu_+$ comb components as a function of negative pump frequency detuning $\delta\nu_-$. %
    $\Delta\nu_\mathrm{ceo}$ is observed to vanish over a finite range, signaling the formation of a single frequency grid thanks to the self-alignment of the PDCS to the pumps. %
    \textbf{d,} Repetition rate disciplining: once synchronized, $\nu_\mathrm{rep}^\mathrm{pdcs}$ directly follows the optical frequency division set by the pump separation ($N = 263$) and is linearly disciplined against the pump frequency detuning. %
    \textbf{e,} Therefore, the synchronization substantially reduces the repetition rate noise compared to the unsynchronized PDCS, detailed in the next section. %
    The power is referenced to \qty{1}{\mW}, i.e. equivalent to \qty{}{\dBm}.
  }
\end{figure*}

\begin{figure*}[!t]
  \centering
  \includegraphics{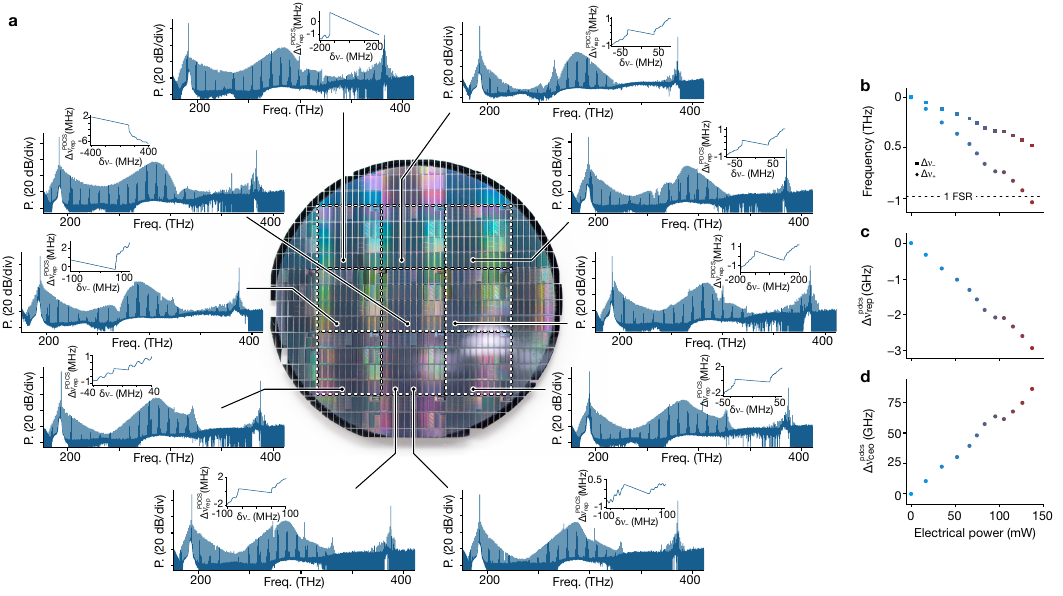}
  \caption{\label{fig:3} 
    \textbf{Wafer-scale self-aligned PDCS yield and active thermal tuning}---
      \textbf{a} The \qty{100}{\mm} silicon nitride wafer, with dashed rectangles indicating the repeating \qty{22}{\mm}$\times$\qty{22}{\mm} reticle field. %
      For 10 devices sampled across the wafer with purposely different coupler designs, all operated at a quasi-harmonic pump setting enabling CEO detection before the chip, the corresponding self-aligned PDCS optical spectrum is shown alongside the repetition rate variation against $\delta\nu_-$ (inset), confirming self-alignment and robustness against coupling variations, thickness variations, and wafer-scale fabrication imperfections, with the required $\nu_-$ adjustment between devices remaining below \qty{\approx130}{\GHz}, well within the heater thermo-optic tuning range. %
      Additional examples of successful operation with different devices are described in Supplementary Information. 
      P: power in dB. %
      \textbf{b-d} Active thermal tuning of self-aligned PDCS comb using integrated heaters (see \cref{fig:1}b). %
      \textbf{b} Joule heating enables continuous tuning of the negative ($\nu_-$, squares) and positive ($\nu_+$, circles) pump frequencies required for self-aligned operation, with tuning of $\nu_+$ over one full free spectral range (FSR) achieved with \qty{135}{\mW} of electrical power. %
      \textbf{c-d} Thermal tuning also enables tuning of $\nu_\mathrm{rep}^\mathrm{pdcs}$ and $\nu_\mathrm{ceo}^\mathrm{pdcs}$, both of which are fully defined by the pump frequencies. %
      In \textbf{b--d}, the reference values are $\nu_- = \qty{192.723}{\THz}$, $\nu_+ = \qty{385.275}{\THz}$, $\nu_\mathrm{rep}^\mathrm{pdcs} = \qty{997.681}{\GHz}$, and $\nu_\mathrm{ceo}^\mathrm{pdcs} = \qty{170.429}{\GHz}$. %
  }
\end{figure*}

\begin{figure*}[!t]
  \centering
  \includegraphics{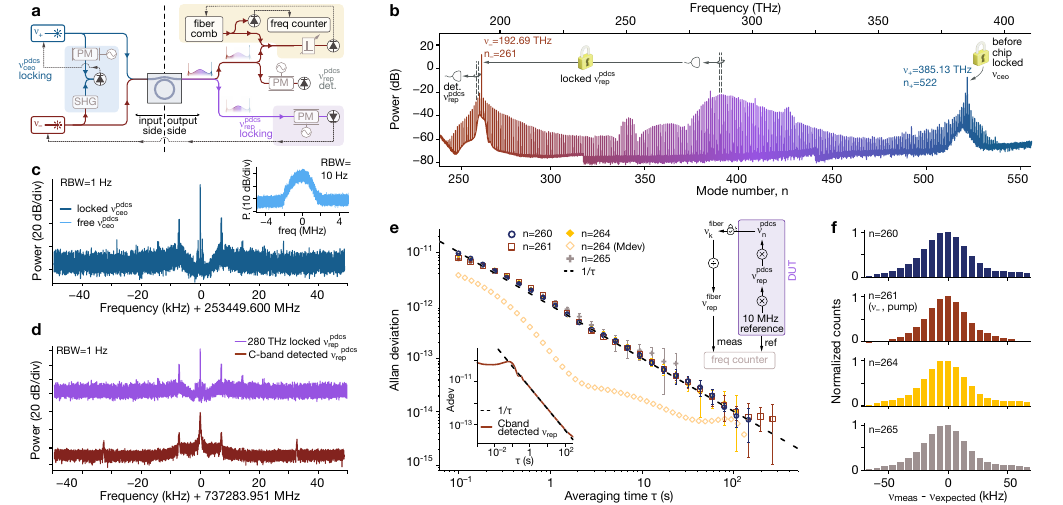}
  \caption{\label{fig:4} \textbf{Microwave-to-optical link: self-aligned PDCS self-referencing for optical frequency synthesis}---
    \textbf{a} Simplified experimental setup for optical frequency synthesis using a self-aligned PDCS microcomb. %
    The CEO locking module (blue), entirely before the chip, doubles $\nu_-$ and beats it against $\nu_+$, with stabilization achieved by feedback onto $\nu_+$. %
    An electro-optic comb (EOcomb) downconverts the CEO beat, which exceeds the direct photodetection bandwidth. %
    The remaining $\nu_-$ degree of freedom locks the repetition rate $\nu_\mathrm{rep}$, detected in the \qty{280}{\THz} band and C-band (near \qty{190}{\THz}) with additional EOcombs (purple module) . %
    A fiber OFC with one comb tooth locked to a selected PDCS tooth (yellow module) enables optical frequency counting of the PDCS. %
    All locks are referenced to a \qty{10}{\MHz} Rb standard. %
    PM: phase modulator, SHG: second harmonic generation. 
    \textbf{b} PDCS spectrum with harmonic pumps at \qty{n_- = 261}{} and \qty{n_+ = 522}{}. %
    The spectrum is normalized to \qty{1}{\mW}, i.e. equivalent to \qty{}{\dBm}. %
    \textbf{c} Locked CEO beat spectrum referenced to the \qty{10}{\MHz} standard (inset: free-running CEO). %
    RBW: resolution bandwidth. %
    \textbf{d} Locked $\nu_\mathrm{rep}^\mathrm{pdcs}$ detected at \qty{280}{\THz} (top) and in the C-band (bottom), verifying a single repetition rate across the PDCS spectrum. %
    \textbf{e} Allan deviation of the PDCS comb teeth (\qty{n = 260}{} to \qty{265}{}), showing the $1/\tau$ slope of synchronization to the microwave reference. %
Error bars: one standard deviation. %
    Right inset: Schematic of the PDCS acting as a coherent microwave-to-optical link (shaded), measured with a fiber OFC. %
    Left inset: Allan deviation of $\nu_\mathrm{rep}^\mathrm{pdcs}$ in the C-band. %
The deviation at $\tau<10^{-1}$~s stems from the poor clocking of the EOcomb microwave synthesizer . %
    DUT: device under test. %
    \textbf{f} Frequency counting accuracy: measured PDCS comb teeth agree with expected values $\nu_{\mathrm{expected}} = n\nu_\mathrm{rep}^\mathrm{pdcs} + \nu_\mathrm{ceo}^\mathrm{pdcs}$.
  }
\end{figure*}

\begin{figure*}[!t]
  \centering
  \includegraphics{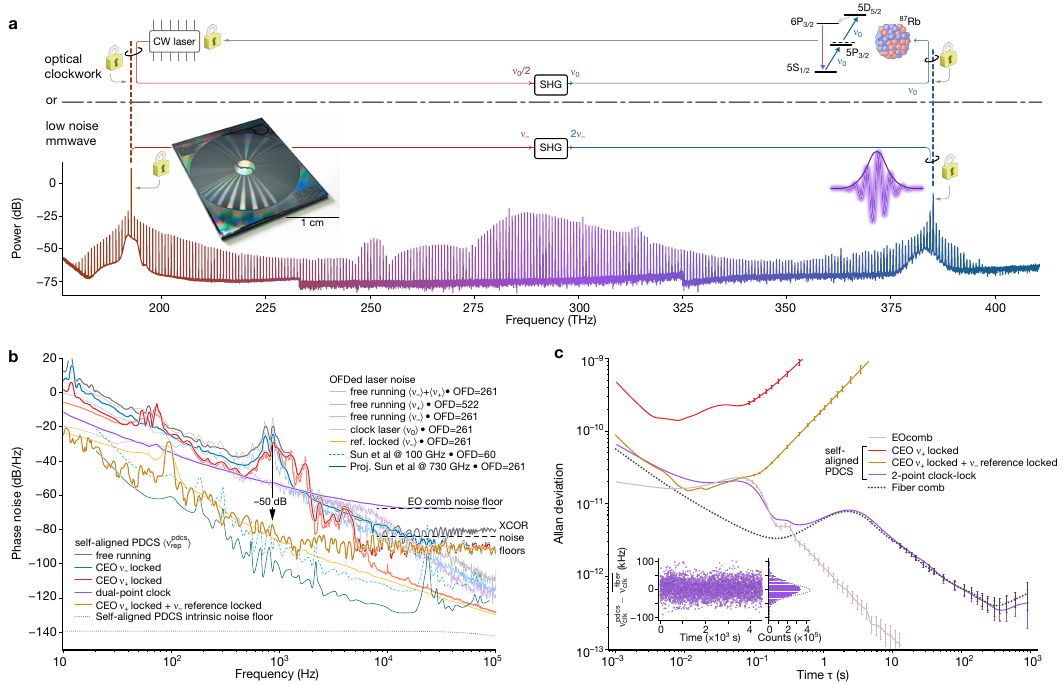}
  \caption{\label{fig:5} \textbf{Optical-to-microwave link: low noise millimeter wave generation and optical clockwork operation}---
    \textbf{a} Experimental system.  %
    Using the same self-aligned PDCS as previously, the system acts as a low-noise millimeter wave generator or as an optical clockwork, depending only on the input laser locks. %
    Power normalized to \qty{1}{\mW} (\qty{}{\dBm}). %
    \textbf{b} PDCS repetition rate phase noise spectral power density $\langle\nu_\mathrm{rep}^\mathrm{pdcs}\rangle$ under different input laser configurations: free running (grey), CEO locked with the negative (blue) or positive pump (red), both lasers locked to the clock laser and its second harmonic (purple), and CEO locked with positive pump and negative pump locked to a reference cavity (yellow). %
    The corresponding laser noise divided by the expected OFD factor appears in matching colors, aligning with the PDCS repetition rate noise. %
    Also shown for comparison are phase noise from Sun et al.~\cite{SunNature2024} (dashed green) and a projection of a self-aligned PDCS in a similar pump-lock configuration.   %
    The self-aligned PDCS photon-lifetime noise floor lies well below the projection (dotted black line). %
    Phase noise densities normalized to \qty{1}{\mW} (dBc/Hz).
    \textbf{c} Allan deviation for different locking configurations. %
Error bars: one standard deviation. %
    The free-running negative pump and CEO locked with positive pump configuration exhibits a rapid random frequency walk (red). %
    The reference cavity locked configuration, despite the best phase noise, drifts over time, leading to poor averaging for $\tau>0.1$ s (yellow). %
    The atomic clock locked PDCS demonstrates good long-term averaging (purple), matching for $\tau>1$ s a commercial fiber OFC clockwork probing the same atomic transition (black dotted). %
    The discrepancy between the fiber comb and PDCS at \qty{\tau<0.1}{\s} originates from the EOcomb $\nu_\mathrm{rep}^\mathrm{pdcs}$ downconverter (grey). %
    Bottom inset: Clock frequency counted from the PDCS and fiber comb repetition rates, $\nu_\mathrm{clk}^\mathrm{pdcs}$ and $\nu_\mathrm{clk}^\mathrm{fiber}$. %
    Both produce identical statistics, shown by the centered Gaussian distribution of $\nu_\mathrm{clk}^\mathrm{pdcs}$ against the fiber comb average $\overline{\nu_\mathrm{clk}^\mathrm{fiber}}$.
  }
\end{figure*} 

We demonstrate the above concept experimentally using silicon nitride (\ce{Si_3N_4}) microring resonators obtained via commercial four-inch wafer foundry fabrication, with a bus waveguide designed to couple across more than an octave (\cref{fig:1}b; see Methods). %
The microring dimensions are carefully designed to engineer the resonator dispersion (\cref{fig:2}a) for both PDCS existence through phase- and frequency-matching and self-alignment to octave-separated pump lasers. %
The first requirement is the degenerate optical parametric oscillation (OPO) condition, minimizing the frequency mismatch $\Delta\nu(n)$ between the pumped and generated resonances at the center mode $n_0$ (see Methods). %
The second, for the self-alignment, requires the PDCS integrated dispersion $D_\mathrm{int}$ to be close to zero at both pumps, minimizing the phase velocity mismatch between the PDCS and its pumps (see Methods). %

\Cref{fig:2} shows octave-spanning, self-aligned PDCS generation. %
The dispersion conditions are satisfied at \qty{\nu_- \approx 193}{\THz} ($n_-=263$, C-band) and \qty{\nu_+ \approx 387}{\THz} ($n_+=2n_-=526$, visible) (\cref{fig:2}a). %
In the linear regime, the positive frequency mismatch $\Delta\nu(n_0)>0$ is sufficiently small that dual-pump thermo-optic effects can achieve $\Delta\nu(n_0)\approx0$ (see Supplementary Information S.3). %
Using on-chip pump powers of \qty{P_-=145}{\mW} and \qty{P_+=70}{\mW} for the pumps at $\nu_-$ and $\nu_+$, respectively, we deterministically generate an octave-spanning PDCS (\cref{fig:2}b; see Methods). %
This procedure produces a continuous single frequency grid, with all comb teeth between the two octave-separated pump lasers above \qty{1}{\nW} (i.e., \qty{-60}{\dBm}). %
Together with the wafer-scale fabrication by which the devices are created and latitude on geometric parameters as shown in Supplementary Information~S.6, we find the realization of self-aligned octave-spanning PDCSs to be as simple as conventional octave-spanning DKSs. 

We demonstrate self-alignment by directly measuring $\Delta\nu_\mathrm{ceo}$ between the PDCS and $\nu_+$ pump comb (see Methods and Supplementary Information S.5). %
Tuning $\nu_-$ actuates $\Delta\nu_\mathrm{ceo}$ (\cref{fig:2}c), which vanishes and locks to zero over a finite pump-frequency range. %
This locking confirms that the PDCS has self-aligned to the pumps as a single octave-spanning comb. %

In a self-aligned PDCS, both pumps are comb teeth that pin $\nu_\mathrm{rep}^\mathrm{pdcs}=(\nu_+-\nu_-)/(n_+-n_-)$. %
The repetition rate therefore depends linearly on the pump detunings $\delta\nu_\pm$, with opposite signs for positive- and negative-pump actuation and slope magnitude $1/N=1/(n_+-n_-)=1/n_-$ corresponding to the optical frequency division (OFD) factor (\cref{fig:2}d). %
Due to self-alignment, the repetition rate noise depends only on the pump frequency noises, reducing its \qty{3}{\dB} linewidth from \qty{70}{\kHz} for the free-running case with three comb components to \qty{1.8}{\kHz} (\cref{fig:2}e). %
  While the conditions required to obtain self-alignment of a PDCS may appear stringent, coupling engineering provides sufficient pump-detuning flexibility to fulfill them across multiple devices and pumped mode pairs (see Methods and Supplementary Information S.6).  %
  This robustness suits wafer-scale fabrication, where unavoidable \ce{Si3N4} film-thickness variations from deposition stress~\cite{OuOpt.Lett.OL2025} shift dispersion, resonance frequencies, and coupling conditions. %
  To demonstrate wafer-scale compatibility, we used another wafer that allows for tests across most of its area. %
  Due to the thicker \ce{Si3N4} deposition, we consider a different nominal geometry from \cref{fig:2}, yielding a different FSR  (see Methods). %
  We show that regardless of the position on the \qty{100}{\mm} wafer, the sampled devices all yield quasi-harmonic self-aligned PDCSs (\cref{fig:3}a). %
  Beyond the 10 devices in \cref{fig:3}a, additional chips of the same nominal design from this wafer and other wafers of the same nominal thickness also generate self-aligned PDCSs, confirming reproducibility across wafer positions and between wafers (see Supplementary Information S.6). %
  The variations in the optical comb spectra in~\cref{fig:3}a reflect the purposely different coupler designs, further demonstrating robustness against coupling variations. %
  Microrings of identical nominal geometry, pumped on the same modes, all produce a self-aligned PDCS, but \ce{Si3N4} thickness variation slightly shifts the absolute pump frequencies required from device to device. %
  Such variations can be compensated by using integrated heaters (see~\cref{fig:1}b); our measurements show that Joule heating enables tuning of $\nu_+$ by more than one $\nu_\mathrm{rep}^\mathrm{pdcs}$ with only \qty{135}{\mW} of electrical power (\cref{fig:3}b), along with $\nu_\mathrm{ceo}^\mathrm{pdcs}$ and $\nu_\mathrm{rep}^\mathrm{pdcs}$ (\cref{fig:3}c,d), all while preserving the self-aligned state (see Supplementary Information S.7). %
  These results establish the foundry compatibility of the platform, for which integrated microcomb CEO detection now becomes a trivial task. %

\section*{Self-aligned PDCS metrology: microwave to optical link and vice versa}  
To demonstrate the potential of the self-aligned PDCS OFC, we show its straightforward implementation in metrology applications. %
As described above, the octave-separated input lasers define a locked CEO frequency that the generated PDCS must possess. %
Self-stabilization of the OFC becomes equivalent to locking the two pumps to each other (through second harmonic generation (SHG)), while another degree of freedom remains available for optical or repetition rate locking. %
Thus, rather than a dedicated DKS for a specific application, our integrated self-aligned PDCS operates bidirectionally between the microwave and optical domains, simply by changing the input lasers' locking point. %
Though standard for commercial fiber-based frequency combs, such operation has not been readily available in the past using microcombs.  %

First, we demonstrate self-stabilization of our PDCS, where both $\nu_\mathrm{rep}^\mathrm{pdcs}$ and $\nu_\mathrm{ceo}^\mathrm{pdcs}$ are locked to a microwave standard, enabling precise knowledge of all optical frequency comb teeth. %
We set the pump lasers at harmonic comb tooth modes (\qty{n_+=2n_-=522}{})  and measure the CEO before the chip from the tapped pumps, with an EOcomb downconverting the large $\nu_+ - 2\nu_-$ beat (\cref{fig:4}a; see Methods and Supplementary Information S.10). After generating the self-aligned PDCS, we lock the downconverted $\nu_\mathrm{ceo}^\mathrm{pdcs}$ to a \qty{10}{\MHz} rubidium microwave standard (\cref{fig:4}c inset) through feedback onto $\nu_+$. %
At the output, we measure $\nu_\mathrm{rep}^\mathrm{pdcs}$ in the \qty{280}{\THz} band with another EOcomb downconverter apparatus (Supplementary Information S.10) and lock it to the same \qty{10}{\MHz} reference (\cref{fig:4}d), feeding back onto $\nu_-$. %

We verify that the self-aligned PDCS is stabilized to the microwave reference by confirming that $\nu_\mathrm{rep}^\mathrm{pdcs}$ is consistent across the OFC spectrum and that all comb teeth carry the same relative stability as the reference. %
First, we measure $\nu_\mathrm{rep}^\mathrm{pdcs}$  in the C-band (see Methods) and confirm the same $\nu_\mathrm{rep}^\mathrm{pdcs}$ \qty{130}{} comb teeth away from the lock point (\cref{fig:4}d). %
The detected $\nu_\mathrm{rep}^\mathrm{pdcs}$ synchronizes to the microwave reference for long-term stability (\cref{fig:4}e inset), despite short-term discrepancies that originate from the EOcomb downconversion apparatus, confirming the self-aligned PDCS has a single repetition rate pinned by the two pump lasers. %
We measure the long-term stability of the self-aligned PDCS comb teeth from \qty{n=260}{} to \qty{n=265}{} using a method akin to stability transfer~\cite{LucasNatCommun2020} from the PDCS to a commercial fiber OFC (see Methods). %
All selected PDCS comb teeth exhibit a $1/\tau$ Allan deviation slope (\cref{fig:4}e), demonstrating the coherent link between the microwave and optical domains, and synchronization to the microwave reference. %
The modified Allan deviation shows a $1/\tau^{3/2}$ slope, confirming white phase noise and synchronization. %
We note that acquisition for the $n=265$ comb tooth is shorter due to limited power available from the fiber-comb apparatus, resulting in a poorer long-term lock.
Finally, we count the optical frequency of the comb teeth (\cref{fig:4}f), which exhibit a Gaussian distribution centered around the expected value from the PDCS locked values of $\nu_\mathrm{rep}^\mathrm{pdcs}$ and $\nu_\mathrm{ceo}^\mathrm{pdcs}$. %

Next we demonstrate octave-separated optical two-point locking that enables coherent OFD from two locked optical tones onto the repetition rate of the self-aligned PDCS, which, with the same architecture, enables either low-noise millimeter-wave generation or optical clockwork operation (\cref{fig:5}). %
Unlike narrow-band two-point locking schemes that typically pin two comb teeth separated by \qty{\approx60}{} teeth through feedback to a single degree of freedom (the single pump)~\cite{PappOptica2014,KudelinNature2024}, our system provides two orthogonal degrees of freedom via the pump lasers, enabling complete stabilization of the repetition rate at the input side. %
Additionally, the octave separation maximizes the OFD factor to $n_+=522$ by incorporating the CEO into the two-point locking scheme, while simultaneously enabling long-term stabilization through clockwork operation and locking to a stable atomic transition within the same architecture. %
Hence, we operate with the same self-aligned PDCS as previously in \cref{fig:4}, changing only the input laser lock point (\cref{fig:5}a). %
In one case, before the chip, the CEO is locked similarly to the previous experiment through feedback onto one pump laser, while now the other remaining pump laser is locked to an integrated \ce{Si3N4} cavity reference~\cite{LiuOptica2022} (see Methods). %
In the other, both pump lasers are locked to a continuous wave laser and its second harmonic, with the latter being locked to the \ce{^{87}Rb} two-photon clock transition\cite{MartinPhys.Rev.Appl.2018} (see Methods). %

We study the phase noise spectral density of the PDCS repetition rate $\langle\nu_\mathrm{rep}^\mathrm{pdcs}\rangle$. %
The PDCS divides down the laser noise ($\langle\nu_-\rangle$, $\langle\nu_+\rangle$ or both) onto $\langle\nu_\mathrm{rep}^\mathrm{pdcs}\rangle$ according to the dual-pinning configuration (see Methods) (\cref{fig:5}b). %
Once the CEO is locked by feedback to the positive pump laser $\nu_+$ and the remaining negative pump laser $\nu_-$ is locked to a \ce{Si3N4} integrated cavity reference~\cite{LiuOptica2022} to reduce its noise, the transduced $\langle\nu_\mathrm{rep}\rangle$ exhibits the expected \qty{50}{\dB} reduction, demonstrating optical frequency division operation of the self-aligned PDCS. %
Although we use cross-correlation detection of $\nu_\mathrm{rep}^\mathrm{pdcs}$, reducing the detection noise by more than \qty{25}{\dB} (see Methods and Supplementary Information S.9), the lasers' long-term frequency drift prevents extended averaging. %
As a result, the noise floor exceeds the OFD of the laser noise above \qty{10}{\kHz}, which future work will improve. %
Despite this measurement floor, our performance at low Fourier frequencies already matches state-of-the-art microcomb sources~\cite{SunNature2024} using a similar integrated \ce{Si3N4} cavity reference~\cite{LiuOptica2022}, despite our intentionally conservative locking configuration and roughly seven times larger carrier frequency. %
An optimized cavity lock projects superior performance owing to our larger OFD factor, while remaining orders of magnitude above the self-aligned PDCS photon-lifetime noise floor (see~Supplementary Information S.1).

 We next operate the same self-aligned PDCS as an optical clockwork, changing only the input laser to be locked to the \ce{^{87}Rb}-SH-stabilized CW laser, providing for frequency stability transfer~\cite{LucasNat.Photon.2023} from the clock laser and its sub-harmonic onto the self-aligned PDCS. %
We note that the phase noise, which we confirm is transferred from the lasers onto $\nu_\mathrm{rep}^\mathrm{pdcs}$ via OFD, exceeds that of the reference-locked case (Fig.~5b), which follows from a lower signal-to-noise ratio of the weak \ce{^{87}Rb} two-photon fluorescence detection~\cite{BeardOpt.ExpressOE2024} compared to a Pound-Drever-Hall lock on our near-critically-coupled cavity~\cite{LiuOptica2022}, decreasing the low-noise performance of the locked laser~\cite{LudlowRev.Mod.Phys.2015}. %
Thus, in our system the atomic reference provides the long-term stability, and the cavity reference the short-term stability.
To that end, we note that in both the free-running and cavity reference-locked self-aligned PDCS cases, the Allan deviation rises rapidly after short averaging time $\tau$, in the former case from intrinsic laser drift while in the latter case because of the thermal drift of the integrated reference cavity~\cite{LiuOptica2022}. %
In contrast, it averages down in the clock-locked case (\cref{fig:5}c) because of the good long term stability of the atomic transition~\cite{MartinPhys.Rev.Appl.2018}. %
To confirm that the self-aligned PDCS acts as a true clockwork, we perform a second experiment a day later in which we locked a fiber comb's CEO and one comb tooth to the clock laser, achieving identical Allan deviation performance except at \qty{\tau<1}{\s}, where the PDCS repetition rate noise is dominated by the EOcomb downconverter. %
We retrieve the measured clock frequency from both the PDCS ($\nu_\mathrm{clk}^\mathrm{pdcs}$) and fiber comb ($\nu_\mathrm{clk}^\mathrm{fiber}$) clockworks (see Methods) via repetition rate frequency counting. %
The PDCS shows a centered Gaussian frequency counting distribution (\cref{fig:5}c inset) offset from $\nu_\mathrm{clk}^\mathrm{fiber}$ by about \qty{3}{\kHz}, well within the $\approx$~\qty{26}{\kHz} standard deviation for seventy-five minutes of statistics. %
Both clockworks exhibit a similar fixed offset from the accepted literature value of the transition, which we attribute to systematic shifts in the Rb two-photon apparatus (see Methods). %

\section*{Discussion}
Our self-aligned PDCS microcomb operates significantly differently from standard DKS microcombs that rely on outward cascading from a centred single pump laser, which results in weak and noisy signals at the comb edges, thus complicating the CEO detection and locking required for precision metrology. %
In our self-aligned system, quasi-harmonic pumping sets the comb span and CEO of the subsequently generated PDCS before the chip, enabling robust operation. %
Moreover, intrinsic two-point locking from the two input pumps generating and trapping the PDCS enables octave-separated orthogonal control of all integrated OFC degrees of freedom. %
Thus, this novel approach enables versatile, bidirectional reconfiguration of the same PDCS as a coherent bridge between the microwave and optical domains. %
We demonstrate the benefit of this versatility by performing three key tasks for integrated OFC systems: self-referencing to establish a stable frequency grid that can serve as a backbone for optical frequency synthesizers~\cite{SpencerNature2018} or wavemeters~\cite{NiuNatCommun2023}, two-point optical locking (including the CEO) for low-noise millimeter wave generation, and optical clockwork for transferring the relative stability of the \ce{^{87}Rb} atomic clock transition to the PDCS repetition rate. %
  Although detecting our large repetition rate currently requires a phase modulator apparatus for downconversion, this high repetition rate is an integral part of the architecture. %
  For a self-referenced platform, a large repetition rate prevents pump power from being diluted into densely packed comb teeth spread across the entirety of the spectrum, most of which serve no metrological purpose. %
  Pairing a large-repetition-rate comb with a separate narrow-band low-repetition-rate comb~\cite{SpencerNature2018,NewmanOptica2019,BrilesOpt.Lett.2018}, in particular in view of recent advances in power efficient, low repetition rate DKS-based microcomb generation~\cite{SternNature2018,ShenNature2020,HelgasonNat.Photonics2023}, for a second level of optical frequency division~\cite{OuCLEO2026} provides a SWaP-efficient path to field deployment, without the prohibitive pump power that an octave-spanning low-repetition-rate device would demand. %
  Additionally, with the recent improvement on integration of $\chi^{(2)}$ material, integrated EO combs are also becoming a viable option for a low-SWaP, second OFD level~\cite{HeSciAdv2024}. %
  Should fabrication advances eventually enable low-SWaP operation of a single octave-spanning comb at a low repetition rate, the benefits of our self-aligned PDCS concept would translate directly to that platform.

Moving forward, our self-aligned PDCS enables several integration capabilities, many arising from the stabilized, octave-separated pumps that remain available for further use. %
For example, the system is highly suitable for integrated SHG of the negative pump for high signal-to-noise ratio CEO detection. %
Available pump powers in the 10~mW range avoid the need for extremely high normalized conversion efficiency. %
Monolithic integration with photogalvanic-induced SHG in silicon nitride microresonators~\cite{ClementiNatCommun2025} is particularly compelling because it would enable a complete OFC clockwork in a single fabrication process. %
In addition, nonlinear processes involving one or both pumps, whose frequencies are precisely known through PDCS self-stabilization, offer significant potential to generate stable continuous-wave reference tones outside the PDCS spectral window. %
Second- and third-harmonic generation and sum-frequency generation can enable interfacing with atomic systems used in quantum sensing~\cite{SaffmanRev.Mod.Phys.2010,PakPhys.Rev.A2022} and higher-performance optical clocks~\cite{LudlowRev.Mod.Phys.2015,NicholsonNatCommun2015,ZhangNature2024}.  %
Finally, while the PDCS center frequency operates in the anomalous dispersion regime, both pumps reside in the normal dispersion regime. %
With appropriate dispersion engineering and pump-frequency selection, we expect self-aligned PDCS bandwidths to exceed the roughly one-octave limit thus far demonstrated with standard DKSs, and thus anticipate that ultra-broadband optical parametric oscillators recently demonstrated in the same silicon nitride platform~\cite{SunLightSciAppl2024} can be utilized for PDCSs. %
Conversely, we note that a self-aligned PDCS, where the two pumps are no longer quasi-harmonically related in a low-repetition rate resonator, is also physically possible based on numerical modeling (see~Supplementary Information S.1), particularly when leveraging recent advances in dispersion engineering\cite{MoilleCommunPhys2023,LucasNat.Photon.2023,YuanNat.Photon.2023}, thus opening interesting avenues beyond self-referencing. %

In conclusion, our self-aligned PDCS presents a foundry-compatible paradigm shift in integrated frequency metrology by inverting the architectural approach that has confined on-chip frequency combs to laboratory settings. %
Combined with rapid advances in laser~\cite{ProkoshinOptica2024,JiNatCommun2026} and modulator~\cite{HeSciAdv2024} integration, our octave-bounded generation scheme is a critical step in fulfilling the long-standing promise of field-deployable precision optical frequency combs for navigation, quantum technologies, and fundamental physics measurements.

\bibliography{Biblio.bib}

\section*{Methods}

\subsection*{Device design and fabrication} 
\vspace{-1.5em}\noindent
The devices were fabricated at the commercial silicon nitride foundry Ligentec~\cite{nist_disclaimer}, where a \qty{100}{\mm} (4~in) wafer is patterned with a repeated \qty{22}{\mm} $\times$ \qty{22}{\mm} reticle containing the \qty{1.8}{\mm} $\times$ \qty{5.4}{\mm} chip with the microring designs of interest. %
The photonic devices are created through a subtractive process, where the \ce{Si3N4} atop a \ce{SiO2} layer is removed to create the design, and then encapsulated in a top \ce{SiO2} layer. %
  The microring geometry, namely the ring width $RW$ and \ce{Si3N4} thickness $H$, is chosen for optimal dispersion conditions where the optical parametric oscillation condition and phase-matching with the PDCS occur simultaneously at the octave-apart pump frequencies $\nu_\pm$. %
  Specifically, the OPO condition results in minimization of the frequency mismatch $2\Delta\nu(n) = [\nu_\mathrm{res}(n_+) + \nu_\mathrm{res}(n_-)] - \left[\nu_\mathrm{res}(n) + \nu_\mathrm{res}(n_+ + n_- -n)\right]$ at the center mode $2n_0 = n_- + n_+$ ($\pm1$ when $n_-$ is odd), where $\nu_\mathrm{res}$ are the resonance frequencies and $n_\pm$ are the pumped modes. %
The self-alignment condition requires minimization of the phase velocity mismatch between the PDCS and its pumps, achieved when the PDCS integrated dispersion, $D_\mathrm{int}(n-n_0) = \nu_\mathrm{res}(n-n_0) - \left[\frac{\nu_\mathrm{res}(n_+) + \nu_\mathrm{res}(n_-)}{2} + \left( n-n_0 \right)\cdot D_1\left(n_0 \right) \right]$, is close to zero at both pumps, where our definition of $D_\mathrm{int}$ has units of linear and not angular frequency. %
This minimizes the frequency offsets $\Delta\nu_\mathrm{ceo}$, with small residual offsets compensated via pump frequency tuning to reach efficient PDCS self-synchronization. %
These conditions essentially reduce to minimizing odd-order dispersion to obtain symmetric PDCS integrated dispersion that crosses zero at the two octave-separated pumps. %
  A bus waveguide of width $W_\mathrm{wg}$, tapered to \qty{200}{\nm} at the facet for edge coupling to lensed fibers, is wrapped in a pulley configuration~\cite{MoilleOpt.Lett.2019} around the microring with a gap $G$ between the bus and ring. %
  The choice of the pulley length $L_\mathrm{c}$ allows for chromatic dispersion engineering of the coupling quality factor $Q_\mathrm{c}$ across the octave, such that the positive pump and the PDCS center frequencies are critically coupled ($Q_\mathrm{c}\approx Q_0$), while the negative pump is largely overcoupled ($Q_0 \gg Q_\mathrm{c}$). %
See Supplementary Information S.2 for details. %
  The in-plane parameters, such as $G$ and $RW$, are varied throughout the chip to compensate for fabrication imperfections and provide some lithographic tuning of the PDCS properties. %
  Results in \cref{fig:2,fig:4,fig:5} used two different devices, both on the same wafer with a \qty{H=660}{\nm} \ce{Si3N4} film. %
The device ring radius is \qty{RR=31}{\um}, corresponding to an FSR of about \qty{735}{\GHz}, exhibiting intrinsic quality factor $Q_0\approx1.5\times10^6$ and coupling quality factor $Q_c(\nu_+)\approx Q_c((\nu_+ +\nu_-)/2)\approx1.5\times10^6$ and $Q_c(\nu_-)\approx 4\times10^{4}$ (see Supplementary Information S.6). %
  The devices in \cref{fig:2,fig:4,fig:5} are coupled to a \qty{W_\mathrm{wg}=500}{\nm} bus waveguide wrapped around the ring at gap \qty{G=400}{\nm} with a pulley length \qty{L_\mathrm{c}=17}{\um}. %
  The two devices differ only in their ring width: the device of \cref{fig:2} has \qty{RW=850}{\nm}, while the device used in \cref{fig:4,fig:5} has \qty{RW=860}{\nm}. %
  The devices used for \cref{fig:3} share the same microring geometry (\qty{RR=23}{\um}, \qty{RW=850}{\nm}) but are fabricated on a different wafer with a \ce{Si3N4} film thickness of \qty{H=680}{\nm}. %
  The coupling conditions are varied from device to device, with pulley lengths $L_\textrm{c}=$ \qty{15}{}, \qty{17}{}, \qty{18}{}, and \qty{19}{\um}, bus waveguide widths $W_\textrm{wg} =$ \qty{450}{}, \qty{460}{}, and \qty{470}{\nm}, and gaps $G=$ \qty{400}{}, \qty{450}{}, and \qty{500}{\nm}, such that the correct combination yields the appropriate coupling condition at the frequencies of interest, as described above, and yields similar Q factors to the \qty{RR=31}{\um} designs. %
  The different coupling configurations change the coupling chromatic dispersion and hence the spectral extraction of the PDCS across the octave, resulting in different output spectra, while not impacting the generation nor the self-alignment of the PDCS state. %
  Across the sampled devices, the required $\nu_-$ frequency adjustment spans \qty{\approx130}{\GHz} (\qty{\approx150}{\GHz} for $\nu^\mathrm{pdcs}_\mathrm{ceo}$), which the integrated heaters compensate through the thermo-optic shift.

\vspace{-1.5em}
\subsection*{PDCS generation protocol}
\vspace{-1.5em}\noindent
To generate the PDCS, the two pump lasers $\nu_-$ and $\nu_+$ are set to the transverse electric (TE) polarization upon injection on-chip to pump the fundamental mode of the microring resonator (TE\textsubscript{0}). %
First, the positive pump $\nu_+$ is set on resonance; given its near-visible wavelength, it can be directly observed through residual scattering captured by the optical vision apparatus, in addition to transmission spectroscopy. %
The negative pump $\nu_-$ thus remains on the blue-detuned side of its resonance. %
Working with $\nu_+$ first and parking it offers the main advantage that, since it operates at a shorter wavelength, the residual absorption and resulting heat generated by this laser exceed that of $\nu_-$ (which is in the C-band). 
This choice prevents complex nonlinear behavior of the pumps' frequency detunings relative to their resonances, and such dual-pump operation, similar to the auxiliary laser approach used for thermo-optic stabilization in standard DKS generation~\cite{ZhouLightSciAppl2019,ZhangOptica2019}, provides intrinsic thermal compensation for temperature stabilization (see Supplementary Information S.3) and direct adiabatic access to the PDCS state. %
The negative pump is then set close to its resonance, with its piezoelectric element sweeping slowly (\qty{20}{\Hz}) across a span of \qty{\approx40}{\GHz}. %
A fiber wavelength demultiplexer enables probing of the PDCS power without interference from the pump powers (as seen in the \cref{fig:1}b inset). %

Overcoupling of the $\nu_-$ resonance offers two main advantages for PDCS generation: it reduces the thermal impact of the $\nu_-$ laser (relative to the fixed $\nu_+$ laser) and enables very large frequency tuning while remaining on resonance. %
Therefore, any frequency discrepancy in the OPO condition for PDCS generation and/or frequency tuning necessary for self-alignment can be compensated while maintaining $\nu_-$ on resonance. %
This makes PDCS generation extremely robust, as it is simply obtained by adiabatic (\textit{i.e.}, manual tuning of the laser actuator) frequency tuning of $\nu_-$, and once accessed, the PDCS remains stable over time (over a working day in the self-aligned regime), without active feedback to the lensed fibers or pump lasers. %
The Supplementary Information S.4 shows that the PDCS, outside of its self-alignment window, can be maintained for \qty{>2}{\GHz} frequency detuning of $\nu_-$. %

\vspace{-1.5em}
\noindent\subsection*{Nonlinear dynamics data acquisition}%
\vspace{-1.5em}\noindent
Once the PDCS is generated through the protocol described above, we characterize the self-synchronization presented in~\cref{fig:2} and Supplementary Information S.6. %
The resonator operates in a small dispersive regime within the bandwidth of interest between $\nu_-$ and $\nu_+$, where the nonlinearity suffices to lock the group velocity between the soliton and any pumped component~\cite{MoilleNat.Photon.2024}. %
Outside of synchronization, the only degree of freedom within the resonator is the phase velocity mismatch, which leads to frequency interleaving of the PDCS and the pumped components. %
This offset $\Delta\nu_\mathrm{ceo}$ is equal in magnitude but opposite in sign between the PDCS and the $\nu_+$ component and the PDCS and $\nu_-$ component. %
Once synchronized, all phases lock, creating a single frequency grid with a single CEO. %
To determine synchronization, we measure $\Delta\nu_\mathrm{ceo}$ and its vanishing when self-alignment occurs. %
The pumped component and PDCS spectra overlap throughout the spectrum, which becomes apparent when working in a soliton crystal state, where the different components can be segregated based on their interleaving (Supplementary Information S.5). %
We detect $\Delta\nu_\mathrm{ceo}$ by filtering the complete spectrum around \qty{300}{\THz} (\textit{i.e.} \qty{980}{\nm}) using a wavelength demultiplexer (WDM) and a \qty{6}{\GHz} photodiode. %
We reconstruct the spectrogram by measuring the temporal signal of $\Delta\nu_\mathrm{ceo}$ while sweeping $\nu_-$. %
Using the \texttt{scipy.signal} python package, we reconstruct the temporal spectrogram. %
We calibrate the negative pump detuning $\delta\nu_-$ by similarly creating a spectrogram of the heterodyne signal between $\nu_-$ and a \qty{125}{\MHz} fiber optical frequency comb. %
A similar procedure is used for the positive pump detuning $\nu_+$. %
The non-zero phase locking window means that the soliton self-adapts to maintain the synchronization condition while the repetition rate can undergo a frequency shift. %
Therefore, $\nu_\mathrm{rep}^\mathrm{pdcs}$ is also disciplined in the self-aligned PDCS case. %
We measure $\nu_\mathrm{rep}^\mathrm{pdcs}$ using an electro-optic comb (EOcomb) downconversion scheme similar to Ref~\onlinecite{MoilleOptica2025}. %
This consists of two cascaded phase modulators that are overdriven by a microwave synthesizer at $\nu_\mathrm{rf}^\mathrm{rep}$ to create many sidebands. %
By modulating two adjacent PDCS comb teeth around \qty{280}{\THz} so that each generates sidebands that span at least \qty{\nu_\mathrm{rep}^\mathrm{pdcs}\approx735}{\GHz}, the $m$\textsuperscript{th} sideband from the two PDCS comb teeth will be offset from each other by a frequency $\nu_\mathrm{IF}^\mathrm{rep}$ within the photodiode detection bandwidth (\qty{50}{\MHz}). %
The EO comb thus acts as a local oscillator, enabling frequency down-conversion of $\nu_\mathrm{rep}^\mathrm{pdcs}$ into a detectable frequency $\pm\nu_\mathrm{IF}^\mathrm{rep} = \nu_\mathrm{rep}^\mathrm{pdcs} - m \nu_\mathrm{rf}^\mathrm{rep}$. %
The sign is determined by whether the modulated PDCS sidebands cross before the overlap; in our case, they are always set to avoid a crossing (\textit{i.e.}, positive sign). %
Because of the linear relationship between the repetition rate $\nu_\mathrm{rep}^\mathrm{pdcs}$ and its frequency down-converted signal $\nu_\mathrm{IF}^\mathrm{rep}$, we can directly map $\partial\nu_\mathrm{IF}^\mathrm{rep}/{\partial\nu_\pm} = \partial\nu_\mathrm{rep}^\mathrm{pdcs}/{\partial\nu_\pm}$ as we keep the EO comb fixed. %
Similarly to $\Delta\nu_\mathrm{ceo}$, we record the temporal signal of $\nu_\mathrm{IF}^\mathrm{rep}$ while sweeping $\delta\nu_-$ and reconstruct the spectrogram. %

\vspace{-1.5em}
\noindent\subsection*{Microwave-to-optical link experiment}
\vspace{-1.5em}\noindent
Self-stabilization of the PDCS is achieved by phase-locking the repetition rate $\nu_\mathrm{rep}^\mathrm{pdcs}$ and the carrier-envelope offset (CEO) frequency $\nu_\mathrm{ceo}^\mathrm{pdcs}$. %
To set $\nu_\mathrm{ceo}^\mathrm{pdcs}$, we double the negative pump $\nu_\mathrm{-}$ through second-harmonic generation (SHG) before the PDCS chip and beat it against the positive pump $\nu_+$. %
Because the PDCS self-aligns to the two pump lasers, both pumps are comb teeth; performing this measurement on the input side is thus analogous to standard $f-2f$ nonlinear interferometry for CEO retrieval, but is done before the comb is generated, is intrinsically low-noise (i.e., only the two lasers contribute, with no impact from the microring cavity or PDCS intrinsic noise), and exhibits a very large signal-to-noise ratio. %
For setting $\nu_\mathrm{ceo}^\mathrm{pdcs}$, the two lasers are tapped from the input using respective \qty{10}{\percent} couplers (without amplification), and we obtain $P_\mathrm{2\nu-} \approx \qty{10}{\micro\watt}$ and $P_\mathrm{\nu+} \approx \qty{1}{\milli\watt}$. %
Since $\nu_\mathrm{ceo}^\mathrm{pdcs}$ is large, its detection is performed using an electro-optic comb apparatus consisting of one phase modulator acting on the positive pump laser to frequency-downconvert $\nu_\mathrm{ceo}^\mathrm{pdcs}$ to a \qty{10}{\MHz} beat note such that $\nu_\mathrm{IF}^\mathrm{ceo} = \nu_\mathrm{ceo}^\mathrm{pdcs} - p\nu_\mathrm{rf}^\mathrm{ceo}$ ($p$ is the number of electro-optic sidebands). %
We phase-lock $\nu_\mathrm{IF}^\mathrm{ceo}$ to a \ce{Rb} \qty{10}{\MHz} microwave reference, which serves as the input reference for all microwave synthesizers throughout the setup, by feedback to the positive pump laser. %
We measure the frequency-downconverted repetition rate $\nu_\mathrm{IF}^\mathrm{rep}$ in the \qty{280}{\THz} spectral band and lock it to the same \qty{10}{\MHz} reference. %
The same downconversion scheme is replicated with a third EOcomb to independently detect $\nu_\mathrm{rep}^\mathrm{pdcs}$ in the C-band. %
From both locks, we determine the expected PDCS comb tooth frequencies such that $\nu_\mathrm{expected}(n) = n\nu_\mathrm{rep}^\mathrm{pdcs} + \nu_\mathrm{ceo}^\mathrm{pdcs} = n\left(\nu_\mathrm{IF}^\mathrm{rep} + m\nu_\mathrm{rf}^\mathrm{rep}  \right) + \nu_\mathrm{IF}^\mathrm{ceo} + p \nu_\mathrm{rf}^\mathrm{ceo}$. %
Although we employed an electro-optic phase modulator due to the specific device large $\nu_\mathrm{ceo}^{\mathrm{pdcs}}$, our self-aligned PDCS architecture also permits ring radius engineering---a standard approach in octave-spanning systems~\cite{ZangOptLett2024}---which we demonstrate in our platform in~Supplementary Information S.8. %
When taken in combination with the thermal tuning shown in~\cref{fig:3}, this enables $\nu_\mathrm{ceo}^{\mathrm{pdcs}}$ to be brought to a desired value, potentially close to zero. %
\\%
\indent We measure an individual PDCS comb tooth frequency and stability using a commercial fiber optical frequency comb with repetition rate \qty{\nu_\mathrm{rep}^\textrm{fiber}\approx 125}{\MHz}. %
We leverage the coherent two-way bridge between optical and microwave frequencies to count the PDCS comb teeth. %
Although one could self-reference the fiber comb by locking its CEO and repetition rate, and then heterodyne one fiber comb tooth against one PDCS tooth, the noise multiplication from such an upstream scheme produces excessive noise in the fiber comb teeth given the modest noise performance of our Rb 10 MHz microwave reference, rendering such frequency counting impossible. %
Instead, we leverage optical frequency division. %
In this scheme, the down-converted CEO of the PDCS, $\nu_\mathrm{IF}^\mathrm{ceo}$, and the CEO of the fiber comb, $\nu_\mathrm{ceo}^\mathrm{fiber}$, are both locked to the same \ce{Rb} \qty{10}{\MHz} microwave reference. %
One fiber comb tooth is then phase-locked to one PDCS tooth at a frequency offset $\delta_\mathrm{lock}$ referenced to the same \ce{Rb} reference. %
By directly detecting and counting $\nu_\mathrm{rep}^\mathrm{fiber}$, we retrieve the PDCS comb tooth frequency and stability with $\nu_\mathrm{meas}(n) = k \nu_\mathrm{rep}^\mathrm{fiber} + \nu_\mathrm{ceo}^\mathrm{fiber} \pm \delta_\mathrm{lock}$, where $k$ is the fiber comb tooth order that locks to the PDCS, and the sign before the lock offset depends on the mode number. %
The measurements presented in \cref{fig:4} are performed deliberately with different $\nu_\mathrm{ceo}^\mathrm{pdcs}$ and $\nu_\mathrm{rep}^\mathrm{pdcs}$ conditions by actuating the pump laser, demonstrating that the self-aligned PDCS comb teeth are predictable regardless of the self-stabilization operating point. %

\vspace{-1.5em}
\noindent\subsection*{Optical-to-microwave link experiment}
\vspace{-1.5em}\noindent
We use the same self-aligned PDCS as in the microwave-to-optics experiment, but modify the locks of the input lasers. %
In one scenario to generate low-noise millimeter waves, we set the CEO of the PDCS similarly by doubling the input negative laser and beating it against the input positive laser, followed by phase-locking the EO frequency down-converted tone to a \qty{10}{\MHz} reference with feedback on either laser depending on the scheme. %
The remaining laser can either remain free or be locked to a reference cavity. %
When the CEO is locked through $\nu_+$, we use an integrated \qty{10}{\m} coiled-spiral microcavity reference made of \ce{Si3N4} and embedded in \ce{SiO2}, similar to that used in Ref.~\cite{LiuOptica2022}, and packaged with direct fiber pigtail access. %
Using a Pound-Drever-Hall technique, where the input laser $\nu_-$ is phase-modulated at \qty{25}{\MHz}, the laser locks to the dip of the high-Q resonance of the integrated cavity reference. %
For noise measurement, the repetition rate $\nu_\mathrm{rep}^\mathrm{pdcs}$ is detected in the same fashion as before, namely using EO combs as local oscillators to frequency down-convert the repetition rate into a detected window. %
However, as with every local oscillator down-conversion scheme, the noise of the EO comb imprints on the detected beat note such that $\langle\nu_\mathrm{IF}^\mathrm{rep}\rangle = \langle\nu_\mathrm{rep}^\mathrm{pdcs}\rangle + N \langle\nu_\mathrm{rf}^\mathrm{rep}\rangle$. %
To bypass the noise floor from the EO comb detection, we detect $\nu_\mathrm{rep}^\mathrm{pdcs}$ simultaneously in the \qty{280}{\THz} and C-band spectral windows with two independent EO combs and microwave synthesizers. %
The two down-converted beats enable a cross-correlation (XCOR) measurement, removing the uncorrelated noise from each EO comb and decreasing the noise floor by approximately \qty{25}{\dB} compared to single EO comb $\nu_\mathrm{rep}^\mathrm{pdcs}$ detection. %
This noise floor can be further reduced with larger XCOR averaging, which in our setup is currently limited by $\nu_\mathrm{rep}^\mathrm{pdcs}$ frequency drift from the reference cavity frequency drift, as seen in the Allan deviation characterization in \cref{fig:5}c. %
The PDCS repetition rate noise follows the dual-pinning configuration of the input lasers. %
Free running pump lasers lead to $\langle\nu_\mathrm{rep}^\mathrm{pdcs}\rangle = (\langle\nu_-\rangle + \langle\nu_+\rangle)/n_-$. %
Locking the CEO on the negative pump leads to $\langle\nu_\mathrm{rep}^\mathrm{pdcs}\rangle = \langle\nu_+\rangle/n_+$, while locking the CEO on the positive pump leads to $\langle\nu_\mathrm{rep}^\mathrm{pdcs}\rangle = \langle\nu_-\rangle/n_-$. %
It is assumed that the CEO locking noise is negligible since it is further optically frequency divided, $\langle\nu_\mathrm{ceo}^\mathrm{pdcs}\rangle/n_\pm$. \\%
\indent For the clockwork measurement, the clock laser is phase-modulated and stabilized using fluorescence detection from the \ce{^87Rb} vapor cell, analogous to PDH locking but detecting incoherent fluorescence instead of transmitted light, with lock-in amplifier and servo feedback, in an apparatus similar to Ref~\cite{BeardOpt.ExpressOE2024}. %
We retrieve the clock frequency from the PDCS using $\nu_\mathrm{clk}^\mathrm{pdcs} = 2\left(n_-\nu_\mathrm{rep}^\mathrm{pdcs} + \delta_- - \delta_+\right) =2\left[n_-\left( \nu_\mathrm{IF}^\mathrm{rep} + m \nu_\mathrm{rf}^\mathrm{rep} \right) + \delta_- - \delta_+\right]$, where $\delta_-=\nu_- - \nu_0/2=\qty{46334200}{\kHz}$ and $\delta_+=\nu_+ - \nu_0=\qty{-160092120.004}{\kHz}$ are the locked offsets of the two pumps from the clock input. %
We determine these offsets accurately by phase locking and frequency downconversion of the pump-clock beat notes using respective EO combs (see Supplementary Information S.10). %
We measured the repetition rate \qty{\nu_\mathrm{rep}^\mathrm{pdcs} = (737302133.41 \pm 0.05)}{\kHz} over seventy-five minutes, yielding \qty{\nu_\mathrm{clk}^\mathrm{pdcs}=385284566278 \pm 26}{\kHz}. %
We retrieve the fiber comb clock frequency from $\nu_\mathrm{clk}^\mathrm{fiber} = k\nu_\mathrm{rep}^\mathrm{fiber} +\nu_\mathrm{ceo}^\mathrm{fiber} + \delta_k$ with \qty{\delta_k=35}{\MHz} the locked offset of the k\textsuperscript{th} comb tooth from the clock input, \qty{\nu_{ceo}^\mathrm{fiber}=-15}{\MHz}, $k = 3082212$, and the measured \qty{\nu_\mathrm{rep}^\mathrm{fiber}=125002.611}{\kHz}~\qty{\pm~7}{\mHz}. %
This leads to \qty{\nu_\mathrm{clk}^\mathrm{fiber}=(385284566275 \pm 20)}{\kHz}. %
Compared to the published two-photon \ce{^{87}Rb} transition frequency of \qty{(\nu_0=385284566370.400\pm10)}{\kHz}~\cite{NezOpticsCommunications1993,HilicoEur.Phys.J.AP1998,BernardOpticsCommunications2000,EdwardsMetrologia2005,ZhangSciRep2015}{}\cite{NewmanOptica2019}, where the standard deviation represents the average across all the references, the PDCS-determined clock laser frequency exhibits an offset of \qty{\nu_\mathrm{clk}^\mathrm{pdcs} - \nu_0 = (-92 \pm 31)}{\kHz} and the fiber comb determined clock frequency is \qty{\nu_\mathrm{clk}^\mathrm{fiber} - \nu_0 = (-95 \pm 25)}{\kHz}. %
Hence both OFCs exhibit similar offsets, with a discrepancy from each other that is well within their respective standard deviations. %
The observed clock frequency offset is attributed to systematic effects in the vapor cell. %
Vapor density may exceed temperature-based estimates from resistance temperature detectors due to temperature gradients in the compact geometry used in this experiment.%
\ce{^87Rb-^87Rb} collisional shifts, which scale exponentially with temperature, can contribute tens of \qty{}{kHz} of frequency shift. %
Atom-wall collisions likely contribute more significantly in this smaller cell than in larger cells, though these effects remain poorly characterized. %
Importantly, this shift does not compromise operational performance as observed in the Allan deviation in \cref{fig:5}c. %

\section*{\textbf{Data availability}}
\noindent The data that support the plots within this paper and other findings of this study are available from the corresponding authors upon request.

\section*{\textbf{Code availability}}
\noindent The simulation code used for the Lugiato-Lefever and thermal modeling presented in this work is available from the corresponding authors upon request.

\section*{Acknowledgments}
\noindent Any mention of commercial products within this article is for information only; it does not imply recommendation or endorsement by the National Institute of Standards and Technology nor the U.S. Government. %
Approved for public release, distribution is unlimited. %
Public Affairs release approval \#AFRL-2026-0416. %
The views expressed are those of the authors and do not reflect the official policy or position of the U.S. Government. %
We thank Karl D. Nelson of Honeywell Aerospace Technologies for help with fabrication of the silicon nitride coil resonator. %
We thank Saleha Fatema and Zachary Newman for insightful feedback. %

\section*{Fundings}
\noindent G.M., J.S., and K.S. acknowledge partial funding support from the Space Vehicles Directorate of the Air Force Research Laboratory and the NIST-on-a-chip and IMS programs of the National Institute of Standards and Technology. %
G.M., K.S., and M.E. acknowledge partial funding support from Marsden Fund of the Royal Society Te Apārangi [Grant No. 23-UOA-071] and from the Quantum Technologies Aotearoa Programme. %
P.S. and C.M. acknowledge support from collaborative agreement 2022138-142232 and 2023200-142386 with the National Center for Manufacturing Sciences as sub-awards from US DoD cooperative agreements HQ0034-20-2-0007 and HQ0034-24-2-0001 and from NIST contract 60NANB24D106. %
D.J.B. acknowledges funding from the DARPA MTO Gryphon program under award number HR0011-22-2-0008. %

\section*{Author contributions}
\noindent
G.M., M.E., and K.S. conceptualized the project.
G.M., P.S., and M.E. developed the theoretical framework with help from Z.L. and C.M. %
G.M. designed the resonators. %
S-C.O. helped with initial device linear characterization.
M.H., K.L. and D.J.B. designed and fabricated the integrated reference cavity. %
R.B, R.R. and S.K. developed the Rb atomic clock system. %
G.M. led the project, and performed the measurements as single operator, and completed most of the analysis. %
J.S. helped with the metrology analysis. %
G.M., M.E., and K.S. wrote the manuscript, with input from all coauthors.
All the authors contributed to and discussed the content of this paper.

\section*{Competing interests}
\noindent G.M., M.E., and K.S. have submitted a provisional patent application based on aspects of the work presented in this paper. %
The other authors declare no competing interests.

\clearpage

\onecolumngrid
\captionsetup[figure]{name=Fig.}
\setcounter{figure}{0}
\setcounter{equation}{0}
\renewcommand{\thesection}{S.\arabic{section}}
\renewcommand\thefigure{S.\arabic{figure}}    
\numberwithin{equation}{section}
\renewcommand\theequation{S.\arabic{equation}}

\tableofcontents

\pagebreak
\enabletocentries

\section{Numerical investigation of self-aligned PDCSs
\label{supsec:simulations}}
\begin{figure*}[h]
    \centering
    \includegraphics{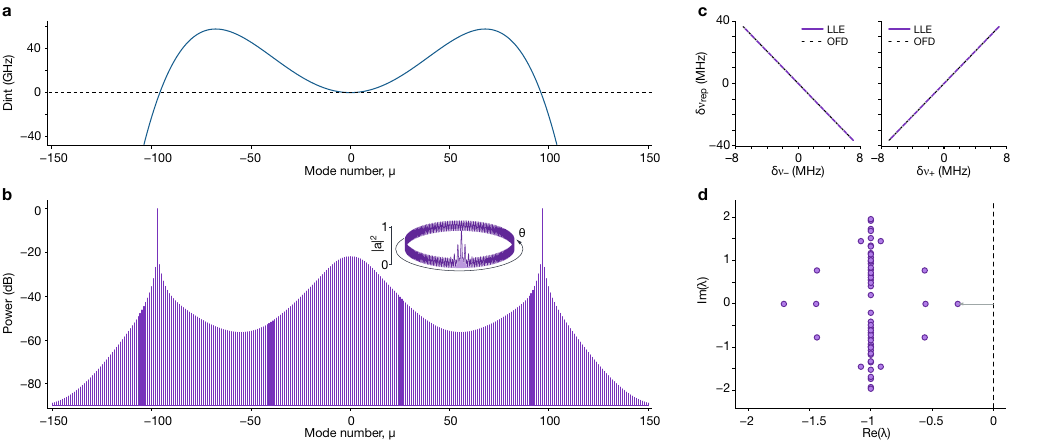}
    \caption{\label{supfig:simulations} \textbf{Numerical study of self-aligned PDCS}---
    \textbf{a} Integrated dispersion profile $D_\mathrm{int}(\mu)$ used in the numerical simulations where only the even order dispersion terms are considered. The dispersion is chosen such that $D_\mathrm{int}(\mu_\pm) \approx 0$ at the pumped modes $\mu_\pm = \pm 97$. %
    \textbf{b} Simulated self-aligned PDCS spectrum. The intracavity field azimuthal profile is shown in the inset. %
    \textbf{c} Repetition rate variation of the PDCS ($\delta\nu_\mathrm{rep}^\mathrm{pdcs}$) as a function of the negative (left) and positive (right) pump detuning variation $\delta\nu_\pm$. The solid lines are the simulation results and the dashed lines the expected disciplining from the optical frequency division between the two pumps. %
    \textbf{d} Eigenvalue spectrum of the linearized operator corresponding to the self-aligned PDCS solution. The arrow represents the position-shifting eigenvalue (Goldstone mode) that has become negative, indicating the symmetry breaking, and hence synchronization to the pumps. 
    }
\end{figure*}

To study numerically the generation and self-alignment of the PDCS, we use the multi-pumped Lugiato-Lefever equation (LLE)~\cite{TaheriEur.Phys.J.D2017}:
\begin{equation}\label{supeq:mlle}
    \begin{split}
        \frac{\partial a(\theta, t)}{\partial t} = -\frac{\kappa}{2}a + i\sum_\mu 2\pi D_\mathrm{int}(\mu)A(\mu) e^{i\mu\theta} + i|a|^2a 
        &+  \sqrt{\kappa_\mathrm{ext}P_\mathrm{-}}\exp\Big\{ i\big[ 2\pi\delta\nu_- + 2\pi D_\mathrm{int}(\mu_-) \big]t + i\mu_-\theta \Big\} \\
    & + \sqrt{\kappa_\mathrm{ext}P_\mathrm{+}}\exp\Big\{ i\big[ 2\pi\delta\nu_+ + 2\pi D_\mathrm{int}(\mu_+) \big]t + i\mu_+\theta \Big\},    
    \end{split}
\end{equation}

\noindent with $a(\theta,t)$ the intracavity field profile along the resonator angle $\theta$ and time $t$, $\kappa$ the cavity loss rate, $D_\mathrm{int}(\mu)$ is the integrated dispersion computed at $(\mu_- + \mu_+)/2$ (i.e., at the spectral center of the PDCS) and has units of frequency (not angular frequency), $P_\pm$ the negative and positive pump power, $\delta\nu_\pm$ the pump detunings from their respective cavity resonances, and $\mu_\pm$ the mode numbers of the pumps relative to the center of the PDCS. The other parameters are defined in the main text.
We chose symmetric conditions with $\mu_+=-\mu_-=97$ to ensure the integrated dispersion $D_\mathrm{int}(\mu_\pm)$ approaches zero at the two pumped modes. %
We account only for even-order dispersion terms, such that $D_\mathrm{int}(\mu) \equiv ({D_2}/{2})\mu^2 + ({D_4}/{24})\mu^4$ (\cref{supfig:simulations}a). %
This approach renders the integrated dispersion and OPO frequency matching condition identical. %

Currently, an understanding of the full phase space of PDCSs and their various excitation pathways remains under exploration~\cite{MoilleSPIEPhotonicsWest2025}. %
Therefore, we assume the single PDCS as an initial condition and allow it to evolve in time, governed by \cref{supeq:mlle}, to confirm that the PDCS is indeed a stable solution of the system. %
From this evolution, we extract the intracavity field profile, consisting of a single pulse with periodic modulation from the beating between the two pumps that travel at the same group and phase velocity, along with its spectrum (\cref{supfig:simulations}b). %
To verify the self-alignment, we proceed similarly to the experiment and scan the negative pump detuning $\delta\nu_-$ while keeping the positive pump detuning $\delta\nu_+$ fixed, and vice versa. %
First, we observe that the PDCS remains stable over the range of detunings studied. %
Second, the PDCS repetition rate variation $\delta\nu_\mathrm{rep}^\mathrm{pdcs}$ changes linearly with the pump detuning variation and the modal separation between the two pumps, as expected from optical frequency division (\cref{supfig:simulations}c). %
We note that the LLE is FSR agnostic, and all the dispersion profiles presented here are as a function of the mode number. %
Therefore, as long as such dispersion can be produced, regardless of the cavity size, a PDCS and more specifically, a self-aligned PDCS, can exist. %

We study the stability of the PDCS by linearizing \cref{supeq:mlle} around the PDCS solution and finding the eigenvalues $\lambda$ of the linearized operator, following the approach in Ref.~\cite{MoilleOptica2025} and described in more detail in Refs.~\cite{MenyukNanophotonics2016,QiOptica2019}. %
We observe that the position-shifting eigenvalue (the Goldstone mode) $\lambda_\mathrm{ps}$ becomes negative, confirming the symmetry breaking and self-alignment of the PDCS (\cref{supfig:simulations}d). %
  From this linear analysis, we can define a Langevin equation, assuming the exponential damping of the fluctuation of the repetition rate: 
  \begin{equation}
    \label{supeq:langevin_dnu}
  \frac{\partial \Delta\nu_\mathrm{rep}^\mathrm{pdcs}}{\partial t} = \frac{1}{2\pi}\left(\lambda_\mathrm{ps} \kappa + \eta_T \frac{d \Delta T}{\partial t}\right)
  \end{equation}
 with $\eta_T$ the variation of repetition rate against temperature and the temperature fluctuation defined by another Langevin equation, 
 \begin{equation}
   \frac{d \Delta T}{\partial t} = -\Gamma_T \Delta T+ \zeta_T 
   \label{supeq:langevin_dT}
 \end{equation}
 where $\Gamma_T$ is the thermal dissipation rate and $\zeta_T$ is a white-noise fluctuation source defined by its autocorrelation,
 \begin{equation}
   \langle \zeta_T(t)\zeta_T(t+\tau)\rangle = \frac{\Gamma_T k_\mathrm{B} T^2}{\rho C V}\delta(\tau)
   \label{supeq:white_noise_fluc}
 \end{equation}
 with $k_\mathrm{B}$ the Boltzmann constant, $T$ the cavity temperature, $\rho$ the \ce{Si3N4} density, $C$ the specific heat, and $\Gamma_T$ the thermal dissipation rate; detailed parameter values for the cavity appear in Refs~\cite{BaiIEEETrans.Ind.Electron.2009,HuszankJRadioanalNuclChem2016,DrakeNat.Photonics2020,MoilleOptica2025}.

Taking the Fourier transform of~\cref{supeq:langevin_dnu}, we obtained the power spectral density of the self-aligned PDCS repetition rate $\nu_\mathrm{rep}^\mathrm{pdcs}$ noise,
 \begin{equation}
   \label{supeq:noise_self-aligned}
   S(f)= 2 \frac{ k_B T\eta_T^2\Gamma_T}{\rho C V\left( f^2 + \Gamma_T^2 \right)} \times \frac{f^2}{f^2 + \lambda_\mathrm{ps}^2 \kappa^2}
 \end{equation}
 The Goldstone mode $\lambda_\mathrm{ps}$ can reach a minimum value of $-1$, establishing a photon-lifetime-limited noise floor for Fourier frequencies $f$ below the photon lifetime inverse ($\kappa$), and the spectrum of \cref{supeq:noise_self-aligned} is displayed in Fig.~5b

\section{Pulley coupler chromatic dispersion
\label{supsec:coupling_design}}

\begin{SCfigure}[][h]
    \centering
    \includegraphics{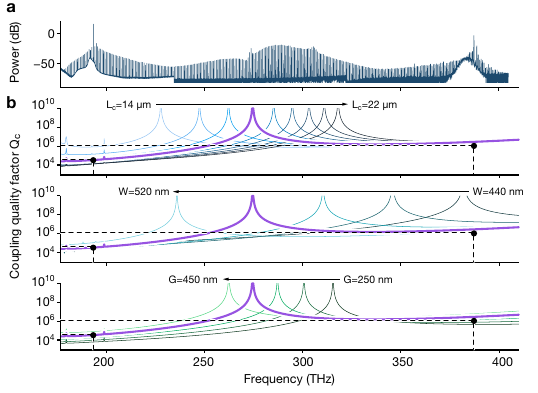}
    \caption{\label{supfig:coupling} \textbf{Pulley-coupler chromatic dispersion engineering for PDCS generation}---
    \textbf{a} Self-aligned PDCS spectra for \qty{RW=850}{\nm} presented in Fig.~1. %
    The pulley coupler characteristics of these devices, and the other $RW$ tested in \cref{supsec:resilience} are \qty{G=400}{\nm}, \qty{L_\mathrm{c} = 17}{\um}, and \qty{W=500}{\nm}, resulting in a coupling quality factor at the pump of $Q_\mathrm{c}(\nu_-) \approx 40\times10^3$ and $Q_\mathrm{c}(\nu_+)\simeq 1.2\times10^6$, which is overcoupled and critically-coupled respectively.%
    \textbf{b} Finite element method simulations of the pulley coupling quality factor against the frequency for a waveguide with the nominal dimensions but varying pulley length (top), varying waveguide width (center), and varying gap size (bottom). The purple trace in the three subplots corresponds to the nominal design used in the experiments. The dashed lines and circles correspond to the two pumps and their respective targeted coupled quality factor.
    } 
\end{SCfigure}

For efficient PDCS generation and self-alignment to octave-apart pump lasers, the chromatic dispersion of the microring coupling to a bus waveguide is critical. %
Although optimization for lowering the degenerate-OPO threshold and optimizing the coupling in every band of interest is still a very active research topic~\cite{LukensOpt.Express2026,TatarinovaPhys.Rev.Appl.2025}, 
we find that for robust self-aligned PDCS generation the two pumps must be simultaneously injected into the microring resonator with high efficiency for effective nonlinear interaction, while providing sufficient frequency tunability to satisfy both the OPO condition and the self-alignment frequency offset condition. %
To achieve this, we design the bus waveguide to couple to the microring resonator in a pulley-like configuration, which exploits the phase mismatch between the bus and the microring resonator to engineer the chromatic dispersion of the coupling quality factor $Q_\mathrm{c}$. %
In all devices tested in this manuscript that yield a self-aligned PDCS (see \cref{supsec:resilience}), the pulley coupler consists of a bus waveguide (with \ce{Si3N4} thickness equal to that of the ring) with a width of \qty{W=500}{\nm}, wrapped around the ring over a pulley length of \qty{L_\mathrm{c}=17}{\um}, and separated by a gap of \qty{G=400}{\nm}. %
This coupler enables over-coupling at the $\nu_-$ pump while maintaining critical coupling at the $\nu_+$ pump, providing greater flexibility in PDCS generation and self-alignment while remaining efficient and allowing for on-chip pump powers approaching or below one hundred milliwatt. %
We note that our two pumps are an octave apart, making the pulley coupler extremely robust against fabrication uncertainty. %
One can first assume that the pulley length remains relatively accurately fabricated. %
A variation of waveguide width of \qty{\pm20}{\nm} results in the same coupling regime at both pumps, while gap variation has limited impact on the coupling (\cref{supfig:coupling}). %

\section{Impact of temperature and thermo-refractive dispersion
\label{supsec:temp}}
\begin{SCfigure}[][h]
    \centering
    \includegraphics{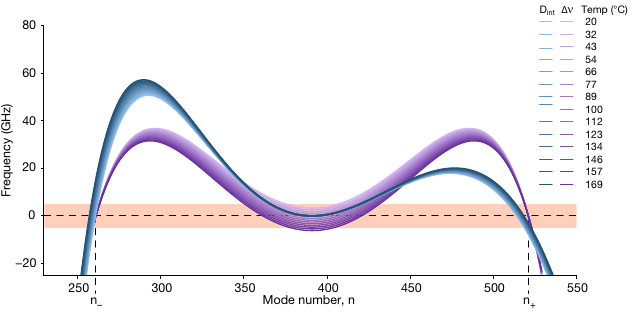}
    \caption{\label{supfig:tempr} \textbf{Temperature dependence of phase matching for PDCS generation and self-alignment}---
    Using finite-element method (FEM) simulations, we compute the OPO frequency mismatch $\Delta\nu$ (varying purple shades) and the integrated dispersion at the PDCS spectral center $D_\mathrm{int}$ (varying blue shades) as a function of resonator temperature. 
    The increase of temperature, happening due to residual absorption of the pumps, leads to a negative shift in $\Delta\nu$, enabling phase-matching for PDCS generation. 
    However, the integrated dispersion $D_\mathrm{int}$ does not change significantly, enabling the PDCS to remain self-aligned over the temperature range studied.
    }
\end{SCfigure}

Here, we discuss the impact of resonator temperature on PDCS generation and self-alignment. %
As with standard DKSs, the on-chip pump powers and residual absorption of the \ce{Si3N4} resonator increase the resonator temperature. %
Although thermal bistability is generally a detriment for stably accessing DKS states with a single pump laser, it does not hinder PDCS generation and self-alignment. %
In particular, we do not implement frequency sweeping faster than the thermal lifetime to counter thermal bistability. %
Although the system is dual-pumped, we do not believe it operates in an identical fashion to laser-cooling and thermo-optic stabilization of microresonators that enable easier DKS access~\cite{ZhangOptica2019,ZhouLightSciAppl2019}. %
Indeed, the negative pump is very largely over-coupled, so its contribution to the thermal load is minimal compared to the positive pump, and no backaction mechanism stabilizes the resonator temperature. %
However, the positive pump is critically coupled and operated near visible wavelengths where \ce{Si3N4} absorption is larger, leading to a significant thermal load that red-shifts the resonator modes. %
In addition, the large frequency span of interest here (from \qty{1550}{\nm} to \qty{775}{\nm}) means that the mode overlap between the \ce{Si3N4} microring resonator core and the \ce{SiO2} cladding---two materials with different thermo-refractive coefficients---varies significantly with wavelength, leading to effective thermo-refractive chromatic dispersion. %
We numerically model the impact of temperature change from room temperature (\qty{20}{\degreeCelsius}) to \qty{169}{\degreeCelsius} on the resonator dispersion using finite-element method (FEM) simulations, extracting the integrated dispersion at the PDCS spectral center $D_\mathrm{int}$ and the OPO frequency mismatch $\Delta\nu$~[\cref{supfig:tempr}]. %
We observe the behavior of $\Delta\nu$ close to the degenerate-OPO mode $(n_-+n_+)/2$, and find that it starts at positive values at room temperature and becomes smaller with increased temperature, eventually crossing zero and becoming negative. %
This indicates that the OPO condition shifts slightly between the cold cavity during linear measurements and the hot cavity when laser pumps are injected. %
This shift explains the slight positive offset of the OPO conditions for the device of interest (Fig.~2a and \cref{supfig:diff_geometry}). %
Importantly, within the temperature range studied, $D_\mathrm{int}$ does not change significantly, and the two pumps remain close to a working point of $D_\mathrm{int}(M_\pm)\approx 0$, enabling the PDCS to become self-aligned. %

One important note is that the over-coupling of the pump resonances in the $\nu_-$ band, and the resulting large linewidths, enables large frequency tuning of the OPO and self-alignment conditions. This makes self-aligned PDCS generation robust in overcoming both fabrication-induced offsets from the target geometry and undesirable thermal effects that might otherwise be a limitation. %

\section{Unsynchronized PDCS repetition rate dependence with pump detuning
\label{supsec:repratePDCS}}
\begin{SCfigure}[][h]
    \centering
    \includegraphics{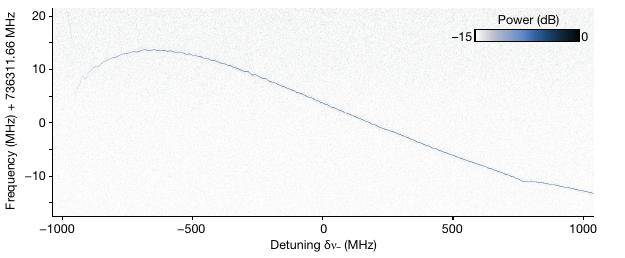}
    \caption{\label{supfig:reprate} \textbf{Repetition rate dependence on the negative pump of the unsynchronized PDCS}--- 
    Repetition rate spectrogram of the unsynchronized PDCS as a function of the negative pump detuning variation ($\delta\nu_-$) across more than \qty{2}{\GHz}.}
\end{SCfigure}
Although we focus the main text on the self-aligned PDCS, we also study the repetition rate dependence of an unsynchronized PDCS with respect to negative pump detuning variation $\delta\nu_-$. %
In this regime, the repetition rate variation $\delta\nu_\mathrm{rep}^\mathrm{pdcs}$ is not expected to follow the optical frequency division relation because the PDCS is not synchronized to the pumps. %
We observe that the repetition rate $\nu_\mathrm{rep}^\mathrm{pdcs}$ increases as $\delta\nu_-$ decreases (\cref{supfig:reprate}), opposite to the self-aligned PDCS case (see Fig.~2). %
Further work is underway to understand this behavior. %
We probe the unsynchronized PDCS over a detuning range exceeding \qty{2}{\GHz} and limit the range only to ensure PDCS stability during measurement. %
This flexibility and robustness against negative pump detuning stems directly from the over-coupling regime of the negative pump resonance, which exhibits a linewidth of nearly \qty{5}{\GHz}. %
We note that the PDCS repetition rate exhibits a turning point near $\delta\nu_- \approx \qty{-600}{\Hz}$, which may prove valuable for future low-noise microwave generation applications and pump noise rejection. %

\section{PDCS-crystal states and PDCS envelope extraction
\label{supsec:multi-DKS}}
\begin{SCfigure}[][!h]
    \centering
    \includegraphics{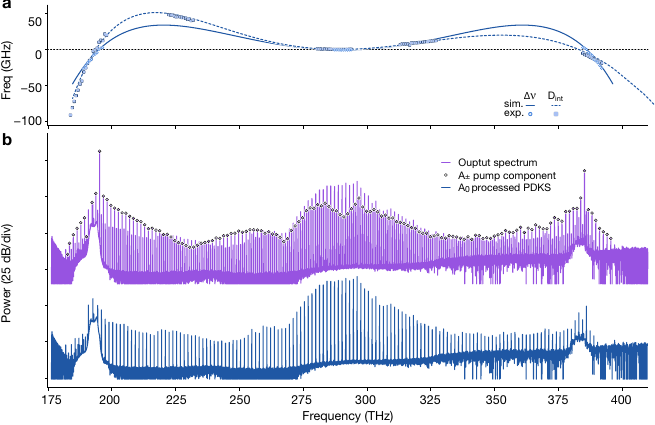}
    \caption{\label{supfig:pdCs_crystal} \textbf{PDCS crystal for post-processing out pump components}---
    \textbf{a} Integrated dispersion $D_\mathrm{int}(\mu)$ (dashed) and OPO frequency mismatch $\Delta\nu$ (solid) for a device with \qty{840}{\nm} ring width. %
    \textbf{b} Under appropriate pump conditions, determined primarily by pump power, a two-PDCS-crystal structure emerges (top purple). %
    In such a scenario, the PDCS crystal and the microcomb generated by cross-phase modulation around the pumps are offset by one mode, in addition to being slightly offset from one another outside the self-alignment region. %
    This configuration enables straightforward extraction of the pump microcomb (top markers), which can be subtracted from the total spectrum to retrieve the PDCS envelope (bottom blue). %
    This procedure reveals the PDCS dispersive wave previously obscured by the pump microcombs and confirms alignment with the phase-matching condition predicted by $D_\mathrm{int}$. %
    }
\end{SCfigure}

Experimentally, we find that our octave-spanning PDCS, under appropriate pump conditions, which we find empirically to be primarily determined by the pump powers, can emerge as a crystal state, where multiple PDCSs are equally azimuthally spaced within the microring resonator~[\cref{supfig:pdCs_crystal}]. %
In particular, we generate a 2-PDCS-crystal state using the \qty{840}{\nm} ring width device by pumping at modes $n_- = 265$ and $n_+ = 523$ (\textit{i.e.}, \qty{195.2}{\THz} and \qty{385.2}{\THz}). %
Since each PDCS can exist with either a $0$ or $\pi$ phase~\cite{EnglebertNat.Photon.2021,MoilleNat.Photon.2024}, we repeat the experiment until the two PDCS are both in phase, observable through a bright peak at $n_0 = (n_- + n_+)/2$. %
However, since $n_0 - n_- = n_+ - n_0$ is odd, the two pumps are spectrally placed at dark teeth of the PDCS crystal spectrum. %
This configuration enables straightforward extraction of the pump microcomb spectrum by selecting every other tooth from the total spectrum (top spectrum and markers in \cref{supfig:pdCs_crystal}b). %
This post-processing capability reveals spectral features previously hidden by the pump microcombs, most notably the PDCS dispersive wave. %
The extracted dispersive wave frequency aligns precisely with the phase-matching condition predicted by the integrated dispersion $D_\mathrm{int}$ (\cref{supfig:pdCs_crystal}a), confirming the theoretical understanding of PDCS generation. %
Beyond fundamental interest, this ability to separate PDCS and pump-generated spectral components through crystal-state engineering provides an additional degree of freedom for tailoring the output spectrum for specific applications, particularly where isolated PDCS spectral coverage is desired without pump laser contributions. %

\section{Robustness of the self-aligned PDCS for wafer-scale compatibility
\label{supsec:resilience}}
\begin{figure}[H]
    \centering
    \includegraphics{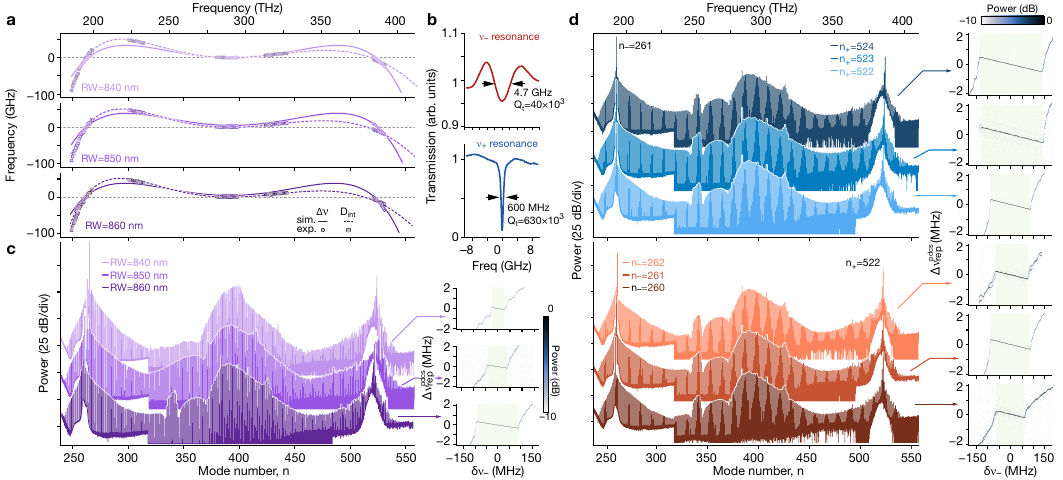}
    \caption{\label{supfig:diff_geometry} \textbf{Fabrication imperfection resilience and pumped mode versatility}---
    \textbf{a} Integrated dispersion $D_\mathrm{int}$ (dashed) and OPO frequency mismatch $\Delta\nu$ (solid) for three different devices with slightly different ring widths (\qty{840}{\nm}, \qty{850}{\nm}, and \qty{860}{\nm}) with experimental data (markers) in agreement with simulation (lines). %
    \textbf{b} Measured pumped resonance spectra for the \qty{860}{\nm} ring width device. The negative pump resonance is strongly overcoupled, with a loaded quality factor $Q_\mathrm{t} \approx 40\times10^3$, while the positive pump resonance is nearly critically coupled, with $Q_\mathrm{t} \approx 640\times10^3$. The overcoupling of the negative pump resonance enables large detuning tolerance for PDCS generation and self-alignment. %
    \textbf{c} Measured PDCS spectra for the three different devices when pumping at harmonic modes. 
    \textbf{d} Measured PDCS spectra for the \qty{860}{\nm} ring width device pumped at different modes for both the positive (top) and negative (bottom) pumps. All the spectrograms shown to the right of the PDCS spectra in (c)-(d) show self-alignment through disciplining of the repetition rate variation $\Delta\nu_\mathrm{rep}^{pdcs}$ against the negative pump detuning $\delta\nu_-$. %
    }
\end{figure}

A critical aspect of our self-aligned PDCS platform is its compatibility with foundry fabrication processes. %
Although foundry-fabricated devices exhibit high quality, they still contain fabrication imperfections that cause deviations from intended design parameters. %
Therefore, resilience to fabrication imperfections constitutes a critical criterion for deployable integrated OFC platforms. %
To assess this resilience, we investigate multiple devices fabricated during the same run. %
We observe that all devices with similar geometry—ring widths from \qty{RW=840}{\nm} to \qty{RW=860}{\nm}, corresponding to approximately \qty{3}{\percent} variation—produce quasi-harmonic self-aligned PDCS with the correct dispersion condition (\cref{supfig:diff_geometry}a-c). %
We implemented additional ring widths in the same fabrication run, but the limited span available for the positive pump (due to the taper amplifier gain bandwidth) prevented testing of all devices. %
In addition, all three tested devices exhibited excellent tolerance against pump mode variation (\cref{supfig:diff_geometry}d), which was again limited only by the available pump laser wavelength range. %
An important contribution to this large tolerance is the chromatic dispersion of the coupling using the pulley coupler~\cite{MoilleOpt.Lett.2019}, which critically couples the positive pump for efficient power transfer to the cavity and leverages a large effective nonlinear coefficient at short wavelengths~\cite{MoilleLaserPhotonicsRev.2025}, while over-coupling the negative pump (\cref{supfig:diff_geometry}b). %
The latter is crucial because it allows the PDCS to exist over a very large negative pump frequency detuning, enabling the system to overcome fabrication imperfections, frequency mismatch, and dispersion offset. %

\begin{figure}[!h]
    \centering
    \includegraphics{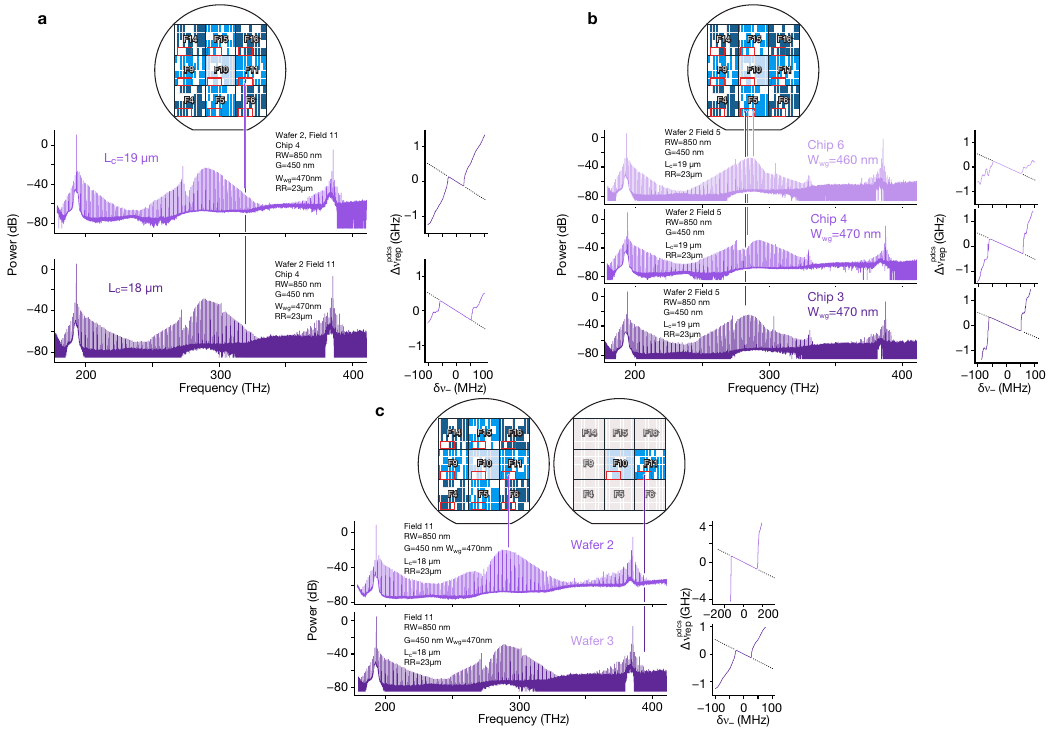}
    \caption{\label{supfig:waver_robust} 
      \textbf{Additional wafer-scale compatibility measurements}---
    All devices use \qty{RW=850}{\nm}, \qty{RR=23}{\um} (\qty{680}{\nm} \ce{Si3N4} thickness, FSR~$\approx$~\qty{1}{\THz}) and are pumped at $\nu_- \approx \qty{192.9}{\THz}$ and $\nu_+ \approx \qty{385.7}{\THz}$. %
    Each panel shows the wafer layout with each field (F) representing a repetitive pattern made of 44 individual chips, of which only the ones highlighted by the red rectangles correspond to microrings for PDCS generation. The blue shade corresponds to the average thickness across the field (light: \qty{677}{\nm}, medium: \qty{680}{\nm} and dark \qty{683}{\nm}). %
    The missing or defective dies are represented in white. For each device tested, we measure their optical spectrum (left) and the corresponding repetition rate disciplining $\delta\nu_\mathrm{rep}^\mathrm{pdcs}$ vs. $\delta\nu_-$ (right), with the dashed line indicating the expected OFD=193 slope. %
    \textbf{a} Same chip (Wafer~2, Field~11, Chip~4), same ring and coupler (\qty{G=450}{\nm}, \qty{W_\mathrm{wg}=470}{\nm}), varying pulley length: \qty{L_\mathrm{c}=19}{\um} (top) and \qty{L_\mathrm{c}=18}{\um} (bottom). %
    \textbf{b} Same field (Wafer~2, Field~5), same ring and pulley length (\qty{L_\mathrm{c}=19}{\um}, \qty{G=450}{\nm}), varying bus waveguide width chip-to-chip: Chip~6 (\qty{W_\mathrm{wg}=460}{\nm}, top), Chip~4 (\qty{W_\mathrm{wg}=470}{\nm}, middle), Chip~3 (\qty{W_\mathrm{wg}=460}{\nm}, bottom). %
    \textbf{c} Same nominal design (Field~11, \qty{G=450}{\nm}, \qty{W_\mathrm{wg}=470}{\nm}, \qty{L_\mathrm{c}=18}{\um}), two fabrication runs: Wafer~2 (top) and Wafer~3 (bottom), both \qty{680}{\nm} thickness.
    }
  \end{figure}

e devices studied above use a microring radius of \qty{RR=31}{\um} (\qty{H=660}{\nm} \ce{Si3N4} thickness), yielding an FSR of approximately \qty{735}{\GHz}, of which only two reticle fields were available, precluding a full-wafer study. %
To extend the robustness assessment to the wafer scale, we fabricated a second wafer with a \ce{Si3N4} thickness of \qty{680}{\nm} and a microring radius of \qty{RR=23}{\um} (\qty{RW=850}{\nm}), yielding an FSR of approximately \qty{1}{\THz}. %
This design change also demonstrates that self-aligned PDCS generation is not restricted to a single FSR. %
For this geometry, all reticle fields across the full \qty{100}{\mm} wafer are available. %
However, not all chips within each field could be characterized: some dies are missing and others are defective (metal residue, damaged facets), and these are highlighted in white in the wafer map of \cref{fig:3}a. %
While the above establishes robustness against ring width and pumped mode variations on a single chip, demonstrating foundry compatibility requires additionally verifying self-alignment across coupling variations, chip-to-chip differences within a field, across independent fabrication runs (\cref{supfig:waver_robust}), and across a full \qty{100}{\mm} wafer (\cref{fig:3}). %
The four complementary experiments presented here systematically isolate these remaining variables. %
In all cases (\cref{fig:3,supfig:waver_robust}), devices are pumped at the same modes around \qty{\nu_-\approx 192.9}{\THz} (\qty{1555}{\nm}) and \qty{\nu_+ \approx 385.7}{\THz} (\qty{777.8}{\nm}). %
In \cref{supfig:waver_robust}a, we test two devices on the same chip (Wafer~2, Field~11, Chip~4) with identical ring geometry (\qty{RW=850}{\nm}, \qty{RR=23}{\um}) and coupler parameters (\qty{G=450}{\nm}, \qty{W_\mathrm{wg}=470}{\nm}), varying only the pulley length: \qty{L_\mathrm{c}=19}{\um} (top) and \qty{L_\mathrm{c}=18}{\um} (bottom). %
Both devices yield self-aligned PDCS, demonstrating robustness against pulley length variation. %
In \cref{supfig:waver_robust}b, we test three chips at different positions within the same reticle field (Wafer~2, Field~5), all sharing the same ring geometry (\qty{RW=850}{\nm}, \qty{RR=23}{\um}) and pulley length (\qty{L_\mathrm{c}=19}{\um}, \qty{G=450}{\nm}), with the bus waveguide width varying chip-to-chip: \qty{W_\mathrm{wg}=460}{\nm} (Chip~6, top), \qty{W_\mathrm{wg}=470}{\nm} (Chip~4, middle), and \qty{W_\mathrm{wg}=460}{\nm} (Chip~3, bottom). %
All three chips yield self-aligned PDCS, confirming within-field reproducibility against chip-to-chip bus waveguide width variation. %
In \cref{supfig:waver_robust}c, we compare devices with identical nominal design (Field~11, \qty{RW=850}{\nm}, \qty{RR=23}{\um}, \qty{G=450}{\nm}, \qty{W_\mathrm{wg}=470}{\nm}, \qty{L_\mathrm{c}=18}{\um}) across two fabrication runs of the same design: Wafer~2 (top) and Wafer~3 (bottom), both with \qty{680}{\nm} \ce{Si3N4} thickness. %
The two devices yield near-identical self-aligned PDCS, confirming run-to-run reproducibility. %
Finally, the full-wafer sweep in \cref{fig:3}a uses a fixed ring geometry (\qty{RW=850}{\nm}, \qty{RR=23}{\um}) throughout, with all devices pumped at approximately \qty{192.9}{\THz} (\qty{1555}{\nm}) and \qty{385.7}{\THz} (\qty{777.8}{\nm}). %
Missing die and chip-to-chip bus waveguide width variations within each field preclude a uniform coupling design across the full wafer; the coupling robustness demonstrated in panels a--c above ensures these unavoidable variations do not prevent self-aligned PDCS generation. %
The combination of these four tests presented in~\cref{supfig:waver_robust} and \cref{fig:3} constitutes a realistic, non-idealized test of foundry compatibility.

\section{Thermal tuning of self-aligned PDCS
\label{supsec:thermal_tuning}}
\begin{figure}[!h]
  \centering
  \includegraphics{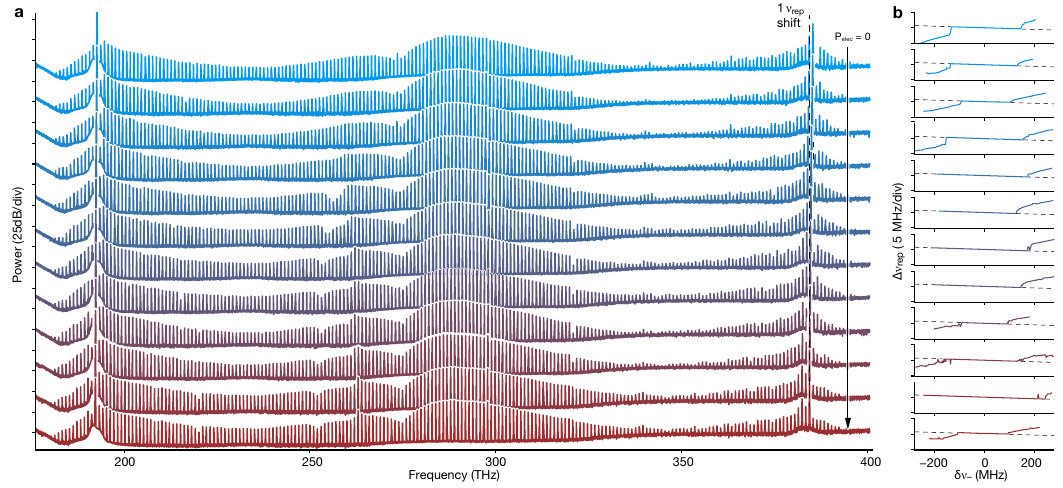}
  \caption{\label{supfig:thermal_tuning}
    \textbf{Thermal tuning of self-aligned PDCS.}~
    \textbf{a,} Optical spectra of the self-aligned PDCS at quasi-harmonic pump settings, recorded at heater powers $P_\mathrm{elec}$ stepped from 0 to \qty{135}{\mW} (blue to dark red), with the integrated heater shown in Fig.~1b of the main text. Colors match those used in Fig.~3b--d of the main text. The positive pump tunes by more than one $\nu_\mathrm{rep}^\mathrm{pdcs}$. %
    \textbf{b,} Repetition rate variation $\Delta\nu_\mathrm{rep}^\mathrm{pdcs}$ as a function of negative pump detuning $\delta\nu_-$ at each power, color-coded to match \textbf{a} and Fig.~3b--d of the main text. The OFD disciplining slope is recovered at every operating point, confirming that self-alignment is maintained across the full thermal tuning range. %
  }
\end{figure}

The integrated heater shown in Fig.~1b of the main text, with a nominal resistance of \qty{\approx 100}{\ohm}, enables thermal tuning of the self-aligned PDCS. %
We step the electrical power $P_\mathrm{elec}$ from 0 to \qty{135}{\mW} in coarse \qty{\approx 10}{\mW} increments, regenerating the PDCS and finding the self-alignment condition at each step. %
Both pumps red-shift via the thermo-optic effect, and since $\nu_+\approx 2\nu_-$, the positive pump tunes by approximately twice the negative pump, covering more than one $\nu_\mathrm{rep}^\mathrm{pdcs}$ (\cref{supfig:thermal_tuning}a). %
The OFD disciplining slope is recovered at each power (\cref{supfig:thermal_tuning}b), confirming self-alignment across the full tuning range. %
A continuous control interface combining fine power stepping with programmed pump-frequency tracking would maintain self-alignment without regeneration, enabling fine alignment of either pump to a reference of interest such as a high-finesse cavity or atomic transition. %

\section{Carrier envelope offset engineering with ring radius
      \label{supsec:ceoVsrr}}
\begin{figure}[!h]
  \centering
  \includegraphics{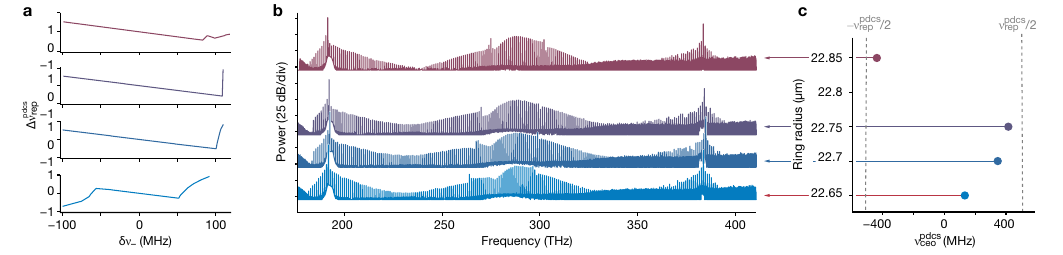}
  \caption{\label{supfig:ceoVsrr}
      \textbf{Ring-radius engineering of the carrier-envelope offset frequency}---
  Colors in \textbf{a} and \textbf{b} correspond to those in \textbf{c}, each identifying a given ring radius. %
  \textbf{a} Repetition rate $\nu_\mathrm{rep}^\mathrm{pdcs}$ as a function of pump detuning $\delta\nu_-$ for self-aligned PDCSs generated in microrings with varying nominal ring radius, demonstrating self-alignment across all ring radii tested. %
  \textbf{b} Optical spectra of the self-aligned PDCS for each ring radius, confirming that the octave-spanning single frequency grid is maintained across the \qty{50}{\nm} step ring radius variation. %
  \textbf{c} CEO frequency $\nu_\mathrm{ceo}^\mathrm{pdcs}$ as a function of ring radius for the corresponding self-aligned states shown in \textbf{b}. %
  }
\end{figure}

The CEO frequency of the self-aligned PDCS is bounded within $[-\nu_\mathrm{rep}^\mathrm{pdcs}/2, \nu_\mathrm{rep}^\mathrm{pdcs}/2]$ and can therefore be large, requiring frequency downconversion for direct detection. %
In standard DKS, ring radius engineering~\cite{ZangOptLett2024} tunes the CEO without altering the dispersion, and has been used to reach directly detectable $\nu_\mathrm{ceo}^\mathrm{pdcs}$~\cite{SpencerNature2018,NewmanOptica2019}. %
The same approach can be applied in the case of self-aligned PDCS. %
To show this, we consider devices from the same wafer as Fig.~3 (\qty{H=680}{\nm} \ce{Si3N4}), with geometry parameters \qty{RW=860}{\nm}, \qty{G=450}{\nm}, \qty{L_\mathrm{c}=17}{\um}, and \qty{W_\mathrm{g}=460}{\nm}. %
For ring radii ranging from \qty{22.65}{\um} to \qty{22.9}{\um}, each device generates a self-aligned PDCS at quasi-harmonic pump frequencies (\cref{supfig:ceoVsrr}a,b), while $\nu_\mathrm{ceo}^\mathrm{pdcs}$ varies by approximately \qty{100}{\GHz} (\cref{supfig:ceoVsrr}c) per device, covering almost the entirety of the CEO space. %
Additionally, this CEO change per \qty{50}{\nm} ring radius variation matches approximately that of the integrated heater tuning shown in Fig.~3d. %
Thus, if the electro-optic phase modulator for CEO downconversion is undesirable, ring radius engineering and thermal tuning together can shift the CEO toward zero, reaching a value directly detectable with off-the-shelf photodetectors.

\section{Cross correlation measurement of the repetition rate
    \label{supsec:xcor_data}
}

\begin{SCfigure}[][h]
    \centering
    \includegraphics{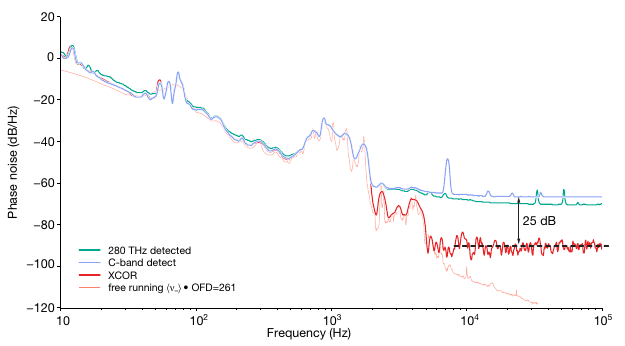}
    \caption{\label{supfig:xcor} \textbf{Cross-correlation measurement of PDCS repetition rate}---
    PDCS Repetition rate phase noise power spectral density with and without cross-correlation (XCOR). %
    Here the positive pump locks the CEO and the negative pump runs freely; hence the repetition rate follows OFD of the latter's noise (pink). %
    When frequency downconverting $\nu_\mathrm{rep}^{\mathrm{pdcs}}$ with an electro-optic phase modulator EO comb apparatus, its noise limits detection beyond \qty{2}{\kHz}. %
    By measuring the repetition rate in both the C-band (blue) and at the center of the PDCS around \qty{280}{\THz} (teal) with independent EO combs, we perform an XCOR measurement where each EO comb downconverter acts as an input to the phase noise analyzer. This rejects the uncorrelated EO comb noise and lowers the noise floor by approximately \qty{25}{\dB} (red). %
    }
\end{SCfigure}

With the repetition rate of the PDCS $\nu_\mathrm{rep}^\mathrm{pdcs}$ being in the \qty{730}{\GHz} range, we cannot directly detect it through photodetection of two adjacent comb teeth. %
Instead, as discussed in the Methods, we use an electro-optic phase modulation apparatus, also known as an electro-optic comb (EO comb), that acts as a local oscillator to down-convert the PDCS repetition rate to a detectable frequency $\nu_\mathrm{IF}^\mathrm{rep} = \nu_\mathrm{rep}^\mathrm{pdcs} - m\nu_\mathrm{rf}^\mathrm{rep}$. %
Therefore, the phase noise power spectral density that we measure is such that $\langle\nu_\mathrm{IF}^\mathrm{rep}\rangle = \langle\nu_\mathrm{rep}^\mathrm{pdcs}\rangle + m \langle\nu_\mathrm{rf}^\mathrm{rep}\rangle$. %
Hence, as in any frequency down-conversion scheme, the local oscillator phase noise $m \langle\nu_\mathrm{rf}^\mathrm{rep}\rangle$ adds to the measured phase noise of the PDCS repetition rate. %
Although high-quality microwave sources are available, the multiplication factor $m$ from the number of EO sidebands needed to span two adjacent PDCS teeth is large ($m \approx 42$ here). %
Therefore, this produces a high noise floor, particularly for Fourier frequencies above \qty{2}{\kHz} (\cref{supfig:xcor}). %

A method to overcome this EO comb noise floor limitation is to use the two input channels of the phase-noise analyzer (PNA) to perform a cross-correlation (XCOR) measurement. %
In such a scheme, the device under test (DUT) is measured in two independent ways, so that the uncorrelated noise from the two DUT inputs is rejected. %
Therefore, by frequency down-converting $\nu_\mathrm{rep}^\mathrm{pdcs}$ using two independent EO combs, their uncorrelated noise will be rejected, leaving only the PDCS repetition rate phase noise. %
We implement such a scheme using the large spectral span of the PDCS and measuring the repetition rate around \qty{280}{\THz} (\textit{i.e.}, close to the PDCS center) and in the C-band, two modes away from the negative pump (see \cref{supfig:setup}). %
Using this technique, the phase noise PSD $\langle\nu_\mathrm{IF}^\mathrm{rep}\rangle$ from both EO comb detections shows the same profile, confirming the same repetition rate throughout the PDCS spectrum. %
Using these two as the DUT inputs of the PNA, we show a reduction of about \qty{25}{\dB} with respect to the single EO comb-limited noise floor, enabling more accurate measurement. %
The limitation in further noise floor reduction comes from the limited averaging available in the XCOR measurement. %
Indeed, the frequency drift, even when locked to the reference cavity (as seen in the Allan deviation in Fig.~4c), limits the total averaging time to about \qty{5}{\min} before the frequency drift prevents proper correlation between the two channels and better averaging. %
However, the currently observed reduction in the noise floor is sufficient to demonstrate the optical frequency division in all the locking regimes studied. %


\section{Complete experimental apparatus
\label{supsec:apparatus}}

\begin{figure*}[h]
    \centering
    \includegraphics{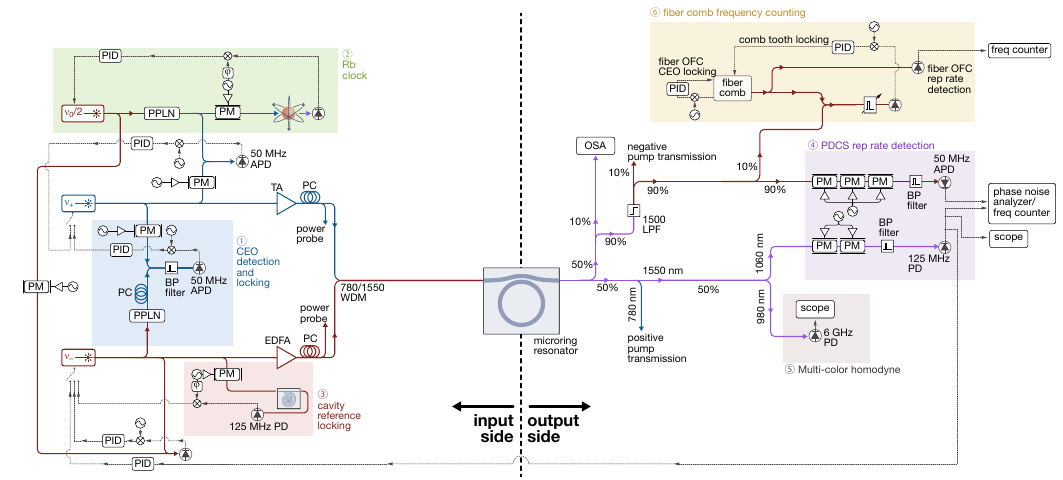}
    \caption{\label{supfig:setup} \textbf{Complete experimental setup}---%
    Complete experimental apparatus used to generate, self-align, and characterize the PDCS. %
    The input side (left) is comprised of the two pump lasers $\nu_-$ and $\nu_+$, amplified by a TA and an EDFA, respectively, that are sent into the microring resonator. %
    Each laser is tapped for setting and stabilizing the CEO of the subsequently generated PDCS comb (blue block \circled{1}), locking the lasers to the Rb clock (green block \circled{2}), or locking the negative pump to the reference cavity (red block \circled{3}). %
    The output side (right) is solely used to characterize the PDCS, with the exception of the self-stabilized microwave-to-optical link case, where the repetition rate detection (purple block \circled{4}) is fed back to the negative pump laser to lock it, and cannot be done otherwise. %
    The fiber comb (yellow block \circled{5}) is used to measure the absolute frequency of the PDCS comb teeth. %
    The multicolor homodyne to detect $\Delta\nu_\mathrm{ceo}$ is displayed in the gray block \circled{6}.
    PC: polarization controller; %
    TA: tapered amplifier; %
    EDFA: erbium-doped fiber amplifier; %
    WDM: wavelength-division multiplexer; %
    OSA: optical spectrum analyzer; %
    PM: electro-optic phase modulator; %
    LPF: long-pass filter; %
    BPF: band-pass filter; %
    PD: photodetector; %
    APD: avalanche photodetector; %
    PID: proportional-integral-derivative controller; %
    PPLN: periodically-poled lithium niobate waveguide; %
    }
\end{figure*}

The complete experimental apparatus used in this work is shown in \cref{supfig:setup}. %
In our self-aligned PDCS system, both pumps generate and trigger the synchronization of the cavity soliton, thereby becoming comb teeth. %
The independent locking of the two pumps on the input side enables stabilization \textit{before PDCS generation}, except in the microwave-to-optical coherent link case, where self-stabilization requires that the detected repetition rate $\nu_\mathrm{rep}^\mathrm{pdcs}$ be fed back for locking. %

Thus, we detect the heterodyne between the doubled negative pump $\nu_-$ and the positive pump $\nu_+$ to lock what will become the CEO frequency $\nu_\mathrm{ceo}^\mathrm{pdcs}$ once the PDCS is generated (\cref{supfig:setup}, block \circled{1}). %
We note that since the CEO detection occurs \textit{before} the chip, the power available exceeds any on-chip light generated (either by the PDCS or any other DKS). %
Therefore, we do not require amplification of the negative pump laser while using a commercial fiber-pigtailed periodically poled lithium niobate (PPLN) crystal and still obtain a doubled power \qty{P_{2\nu-}\approx10}{\uW}, which we beat against a phase-modulated tap of the positive pump, at a power \qty{P_{\nu+}\approx1}{\mW}. %
As described in the Methods section, $\nu_\mathrm{ceo}^\mathrm{pdcs}$ lies beyond our detectable microwave range, and thus the phase-modulation of the tapped pump is done in the overdriven regime, resulting in an electro-optic comb for frequency downconversion. %
The frequency downconverted CEO $\nu_\mathrm{IF}^\mathrm{ceo}$ is then phase-locked to the \qty{10}{\MHz} microwave reference by feedback on one of the lasers, which in \cref{supfig:setup} is the positive laser. %
Alternatively, the input laser can be locked to the Rb clock laser (\cref{supfig:setup}, block \circled{2}) where a CW laser is doubled and locked to the 5S\textsubscript{1/2}-5D\textsubscript{5/2} two-photon transition of the \textsuperscript{87}Rb vapor, which is set in a \qty{100}{\degreeCelsius} gas cell.
The clock laser is phase (frequency) modulated and sent through the vapor cell, similar to a Pound-Drever-Hall (PDH) technique. %
The atoms are preferentially excited by resonant light, and fluoresce as they relax back to the 5S\textsubscript{1/2} state. %
We detect the power of this fluorescence. %
We deliberately phase modulate the laser slower than the atomic relaxation rate in order to keep a good contrast on the detected signal. %
The difference from a standard PDH approach is that the power of the fluorescence (incoherent) is detected instead of the probe laser power, and that there are therefore no off-resonance sidebands for the signal to interfere with. From there it is the same control loop as a PDH, with lock-in and servo feedback. %
The fundamental and second harmonic of this laser are then used to lock the negative and positive pump lasers, respectively, using heterodyne detection and feedback on each laser current. %
Finally, the positive pump can be locked to a high-finesse reference cavity (\cref{supfig:setup}, block \circled{3}) using PDH locking. %

The output of the experimental setup is dedicated to the characterization of the PDCS. %
A portion is extracted to probe the PDCS spectrum using an optical spectrum analyzer (OSA). %
Another portion is sent to the repetition rate detection block (\cref{supfig:setup}, block \circled{4}), using appropriate wavelength demultiplexers (WDMs) to select the desired spectral region for $\nu_\mathrm{rep}^\mathrm{pdcs}$ detection. %
EO combs at \qty{1550}{\nm} (\qty{190}{\THz}) and \qty{1060}{\nm} (\qty{280}{\THz}), consisting of two and three phase modulators respectively, down-convert the PDCS repetition rate to obtain two $\nu_\mathrm{IF}^\mathrm{rep}$ signals. %
In the case of self-stabilization for a microwave-to-optical link, the \qty{1060}{\nm}-detected $\nu_\mathrm{IF}^\mathrm{rep}$ provides feedback to the pump laser to lock the PDCS repetition rate to the \qty{10}{\MHz} reference. %
In the optical-to-microwave link case, the two detected $\nu_\mathrm{IF}^\mathrm{rep}$ are used for the XCOR measurement to reject the EO comb noise floor, as discussed in \cref{supsec:xcor_data}. %
To characterize the optical frequency of the PDCS comb teeth, a portion of the PDCS output is heterodyned against a fiber comb whose CEO is locked to the \qty{10}{\MHz} reference and whose repetition rate is locked to a given PDCS comb tooth. %
This configuration enables precise optical frequency counting from the optical frequency division operation of the fiber comb. %
Finally, a portion of the PDCS is extracted around \qty{980}{\nm} (\textit{i.e.} \qty{305}{\THz}) to monitor the multi-color beat $\Delta\nu_\mathrm{ceo}$ between the PDCS and the pump comb components, which vanishes when the PDCS becomes self-aligned. %

\end{document}